\documentclass[showpacs,floatfix,aps,prb,a4paper,twocolumn]{revtex4}
\usepackage{amsmath}
\usepackage{graphicx}
\usepackage{latexsym,epsfig,bm,times,psfrag,subfigure}
\usepackage{bm,times}
\usepackage{dcolumn}

\begin{document}

\title{Multiparticle interference in electronic Mach-Zehnder interferometers}
\author{D.\ L.\ Kovrizhin$^{1,2}$ and J.\ T.\ Chalker$^{1}$}
\affiliation{$^{1}$Theoretical Physics, Oxford University, 1, Keble Road, Oxford, OX1
3NP, United Kingdom}
\affiliation{$^{2}$RRC Kurchatov Institute, 1 Kurchatov Sq., Moscow, 123182, Russia}
\date{\today}
\pacs{71.10.Pm, 73.23.-b, 73.43.-f, 42.25.Hz}

\begin{abstract}
We study theoretically electronic Mach-Zehnder interferometers built from
integer quantum Hall edge states, showing that the results of recent experiments can be understood in terms of multiparticle interference effects. These experiments probe the visibility of 
Aharonov-Bohm (AB) oscillations in differential conductance as an interferometer is 
driven out of equilibrium by an applied bias, finding a lobe pattern in visibility as a function of voltage.
We calculate the dependence on voltage of the visibility and the phase of AB oscillations at zero temperature, taking into account long range interactions between electrons in the same edge for interferometers operating at a filling fraction $\nu=1$. We obtain an exact solution via bosonization for models in which electrons interact only when they are inside
the interferometer. This solution is non-perturbative in the tunneling
probabilities at quantum point contacts. The results match observations in 
considerable detail provided the transparency of the incoming contact is close to one-half: 
the variation in visibility with bias voltage consists of a series of lobes of decreasing amplitude,
and the phase of the AB-fringes is practically constant inside the lobes but jumps by $\pi$ at the minima of the visibility. We discuss in addition the consequences of approximations made in other recent treatments of this problem. We also formulate perturbation theory in the interaction strength and use this to study the importance of interactions that are not internal to the interferometer.
\end{abstract}

\maketitle

\section{Introduction}\label{sec:introduction}

Recent experiments\cite{heiblum1,heiblum2,roulleau07,litvin07,heiblum3,
neder07, litvin08, roulleau08, bieri08, roulleau09} on electronic
Mach-Zehnder interferometers (MZIs) constructed from integer quantum Hall edge
states have attracted a great deal of attention. In these experiments Aharonov-Bohm (AB) oscillations are observed in the differential conductance of the interferometer. The most striking results concern behaviour at finite bias voltage. The visibility of AB oscillations shows a series of lobes as a function of voltage, while their phase is independent of bias, except near visibility minima where it changes sharply by $\pi$. Our concern in this paper is with the theoretical understanding of these experiments.

The observations are interesting from several perspectives. First, as was quickly appreciated,\cite{heiblum2} it is plausible that the effects arise from electron-electron interactions, because  behaviour  of this kind does not occur in a single particle model. In addition, more seems to be required than a simple treatment in which inelastic scattering leads only to decoherence, since approaches of that kind cannot produce multiple side lobes in visibility of AB oscillations with increasing bias.  It is remarkable that electron interactions should have the distinctive signatures found in this system, since integer quantum Hall edge states are usually modelled in the low energy limit as a chiral Fermi gas of independent particles.\cite{wen} The experiments therefore appear to reflect interaction physics that is not captured by the standard, universal description, but  is robust enough to appear in many devices of varying designs. A second reason for interest stems from current efforts\cite{chamon,goldman,Jonckheere,Law,Kim,feldman06,averin,stern,boyarsky08} to study interferometry in fractional quantum Hall states as a probe of fractional or non-abelian quasiparticle statistics. Against that background it is clearly important to understand unexpected interaction effects in much simpler, integer quantum Hall systems. A third reason for interest is that the phenomenon seems to be an example of coherent many-body physics in a quantum system far from equilibrium. It invites comparison with other non-equilibrium quantum problems, from the Kondo effect\cite{kondo} to cold atomic gases.\cite{atomic}

\begin{figure}[b]
\epsfig{file=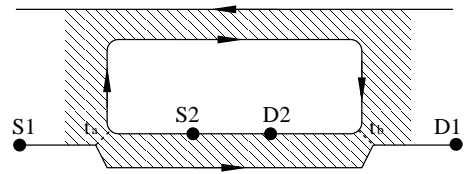,width=6cm}
\caption{Schematic view of the electronic Mach-Zehnder interferometer, which consists of a Hall bar with an island in it. A two-dimensional electron gas in a quantum Hall plateau occupies the shaded region. One edge state propagates along the lower edge of the Hall bar, from source S1 to drain D1, and a second edge state propagates around the island, from source S2 to drain D2.
Tunneling between these two edge states takes place at two quantum point contacts, with amplitudes $t_a$ and $t_b$, at the points indicated by dashed lines.}
\label{fig_machzender_exp}
\end{figure}

The design of an experimental device working as an MZI is shown in 
Fig.~\ref{fig_machzender_exp}. It uses the edge states of a two-dimensional electron gas that is in an integer quantum Hall plateau. (Most experiments have been done at filling factor $\nu{=}2$, but broadly similar results have also been reported at $\nu{=}1$). The edge states serve as electron waveguides and are coupled at quantum point contacts (QPCs), which act as beam splitters. Current between, for example, source S1 and drain D2 is measured as a function of the voltage difference applied between sources S1 and S2. Interference fringes are observed as oscillations in the differential conductance, either when the magnetic flux density is varied by a small amount, or when a side gate is used to change the interferometer area or arm lengths. The visibility and phase of these oscillations vary with voltage in the fashion already summarised. A physical scale is set by the bias voltage at which the first minimum in visibility occurs: the measured value, about 14$\mu$V in the first experiment,\cite{heiblum2} corresponds roughly to the chemical potential increase required to add one electron to an edge state with length equal to that of the interferometer arms: for arm length $d$ and edge velocity $v_F$, this chemical potential increase is   $\hbar v_F/d$.

Theoretical studies of coherence in electronic MZIs started before these experiments. Early work 
treated dephasing arising from a variety of possible sources: interactions within the interferometer; \cite{Seelig01} a fluctuating classical field;\cite{Marquardt04,Forster} voltage probes;\cite{Chung} or coupling to an external quantum bath.\cite{Marquardt06} None of these approaches generates the subsequently observed lobe pattern in the dependence of visibility on bias voltage. A further calculation,\cite{chalker07} based on a microscopic treatment of the effects of long-range interactions and using bosonization combined with a perturbative treatment of tunneling at the QPCs, shows that non-monotonic variations in visibility can appear for weakly coupled edge states, but without capturing the features found experimentally.  By contrast, studies of models with additional structure, involving either a counter-propagating edge mode\cite{sukhorukov07} or the pair of edge modes that arise at filling factor $\nu{=}2$,\cite{sukhorukov08}  show that resonances can appear in that setting, which lead to lobes in visibility similar to those observed. These results are encouraging, but (as we discuss in Section \ref{sec:conclusion}) the models involved seem to us insufficiently generic to account for all experiments. More recently, approximate treatments of the effects of interactions at $\nu{=}1$ when edges are strongly coupled by QPCs have generated some of the behaviour found experimentally.\cite{neder+marquardt,neder08,sim08} There is good reason to think that these calculations identify some of the relevant physics, but the approximations used are non-standard and their domain of validity is unclear.

In this paper we set out a detailed treatment of interaction effects in MZIs at filling factor $\nu={1}$. The approach is microscopic in the sense that it is based on the standard Hamiltonian for quantum Hall edge states,\cite{wen} and does not involve external noise. Our main results come from the exact solution of models which have one simplifying feature: interactions that are restricted to the interior of the MZI; a short account of this part of our work has been presented previously.\cite{exact09} We also present work in three further directions. One of these is an elementary solution of the two-particle problem, which is a simplification of ideas from Ref.~\onlinecite{sim08}. We believe that this calculation provides a useful illustration of the essential physics behind the phenomena we are concerned with, which is multiparticle interference. A second direction is a careful analysis of the approximations involved in Ref.~\onlinecite{neder08}. The third direction is the formulation of perturbation theory in interaction strength, which allows us to assess the importance of interactions that extend beyond the interior of the MZI. 

The organisation of rest of the paper is  as follows. The two-particle problem is addressed in  Section \ref{sec:toy-model}, and the general microscopic description of the MZI is set out in Section \ref{sec:model}.  In Section \ref{sec:exact-solution} we show how models with interactions only in the interior of the interferometer can be solved exactly. In Secton \ref{sec:application_potentials} we use this approach
to study interferometers with various
interaction potentials. We present a extended discussion of Ref.~\onlinecite{neder08} in Section \ref{sec:discussion_neder}, and develop  perturbation theory in interaction strength in Section \ref{sec:pert-theory}. We summarise our conclusions 
in Section \ref{sec:conclusion}. Some technical details of the
calculations are given in appendices.

\section{Two particle problem}\label{sec:toy-model}

In this Section we set out a pedagogical treatment of the two-particle problem that illustrates how electron interactions affect the visibility of AB oscillations in an MZI.
We consider an interferometer having both arms of the same length $d$ and a propagation
velocity $v_F$ for electrons. Denoting their separation by $s$, the flight time during which both are inside the interferometer is $\tau=(d-s)/v_F$. We take the two electrons to interact with a potential energy $U$ when both are inside the MZI on the same edge, but not to interact otherwise. For simplicity, we consider first the case in which the magnitudes of the transmission amplitudes at the two QPCs are $t_a=t_b=1/\sqrt{2}$, giving results for the general case later

We solve the scattering problem for an initial state in which both particles are
positioned on the upper channel before the
first contact. We evaluate the probability for one or both particles to exit the interferometer in the lower channel by summing  all quantum mechanical amplitudes that connect the
initial state to a given final state. We regard the expectation value for the total charge transferred from the upper channel to the lower channel as the analogue for the two-particle problem of the current in the many-body, steady state case.

\begin{figure}[htb]
\epsfig{file=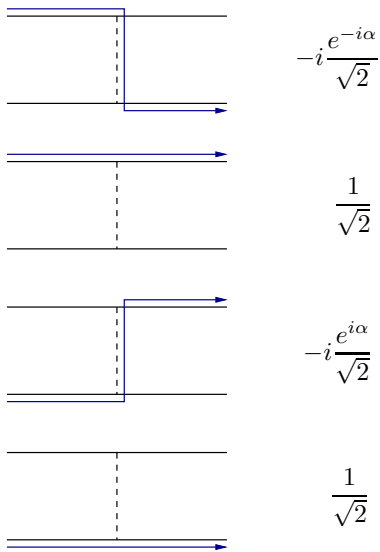,width=4.9cm} 
\caption{Possible paths and associated scattering amplitudes at a single QPC with transmission probability  $1/2$. The phase $\protect\alpha$ of the transmission amplitude is unimportant for a single QPC but contributes to the AB phase in an MZI.
}
\label{fig_scatt_amp}
\end{figure}

\begin{figure}[htb]
\epsfig{file=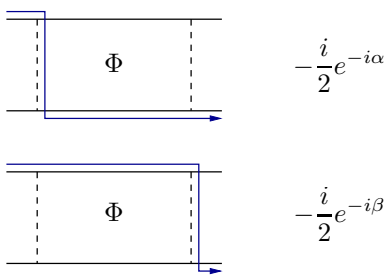,width=5.cm} %\vspace*{-0.5cm}
\caption{Possible paths and associated amplitudes for a single particle to propagate through an
MZI from an initial state on the upper channel to a final state on the
lower channel. The phase difference $\Phi=\protect\beta-\protect\alpha$ is the AB-phase arising from enclosed flux.
 The combined amplitude $A$ for transitions between these states 
is the sum of contributions from the two paths: 
$A=-ie^{-i(\protect\alpha+\protect\beta)/2}\cos(\Phi/2)$.
The total current is proportional to $|A|^2=\frac{1}{2}[1+\cos\Phi]$.}
\label{fig_one_part_paths}
\end{figure}

To establish some notation, consider in the first instance single particle scattering, initially for one QPC and
then for an MZI. Amplitudes for the four scattering processes at one QPC are shown
in Fig.~\ref{fig_scatt_amp}, and the
two possible paths through an MZI between an initial state in the upper channel 
and a final state in the lower channel are shown in 
Fig.~\ref{fig_one_part_paths}. The amplitudes associated with these two  paths
are given by products of the amplitudes arising at each QPC.
Taking the total current to be
proportional to the transition probability between the upper and lower channels, one
obtains the standard result given in the caption to Fig.~\ref{fig_one_part_paths}, with oscillations in the current as a function of the AB phase $\Phi$.

\begin{figure}[hb]
\epsfig{file=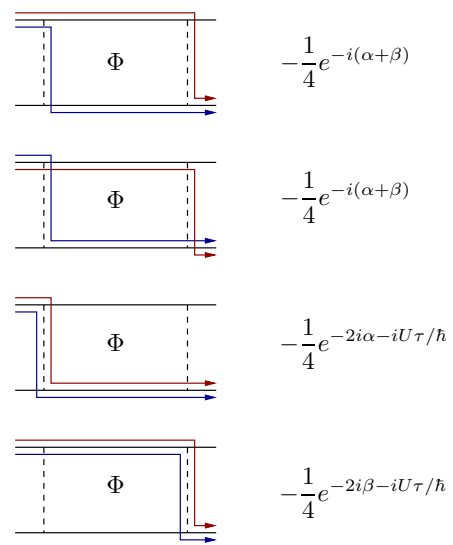,width=5.7cm} 
%\vspace*{-0.5cm}
\caption{(Color online) The four possible paths and associated amplitudes
for two particles to propagate through an MZI from an initial
state with both particles on the upper channel to a final state with both
particles on the lower channel. The combined amplitude is $A_2=-%
\frac{1}{2}e^{-i(\protect\alpha+\protect\beta)}[1+e^{-i U\protect\tau%
/\hbar}\cos\Phi]$.}
\label{fig_two_part_paths_1}
\end{figure}

\begin{figure}[hb]
\epsfig{file=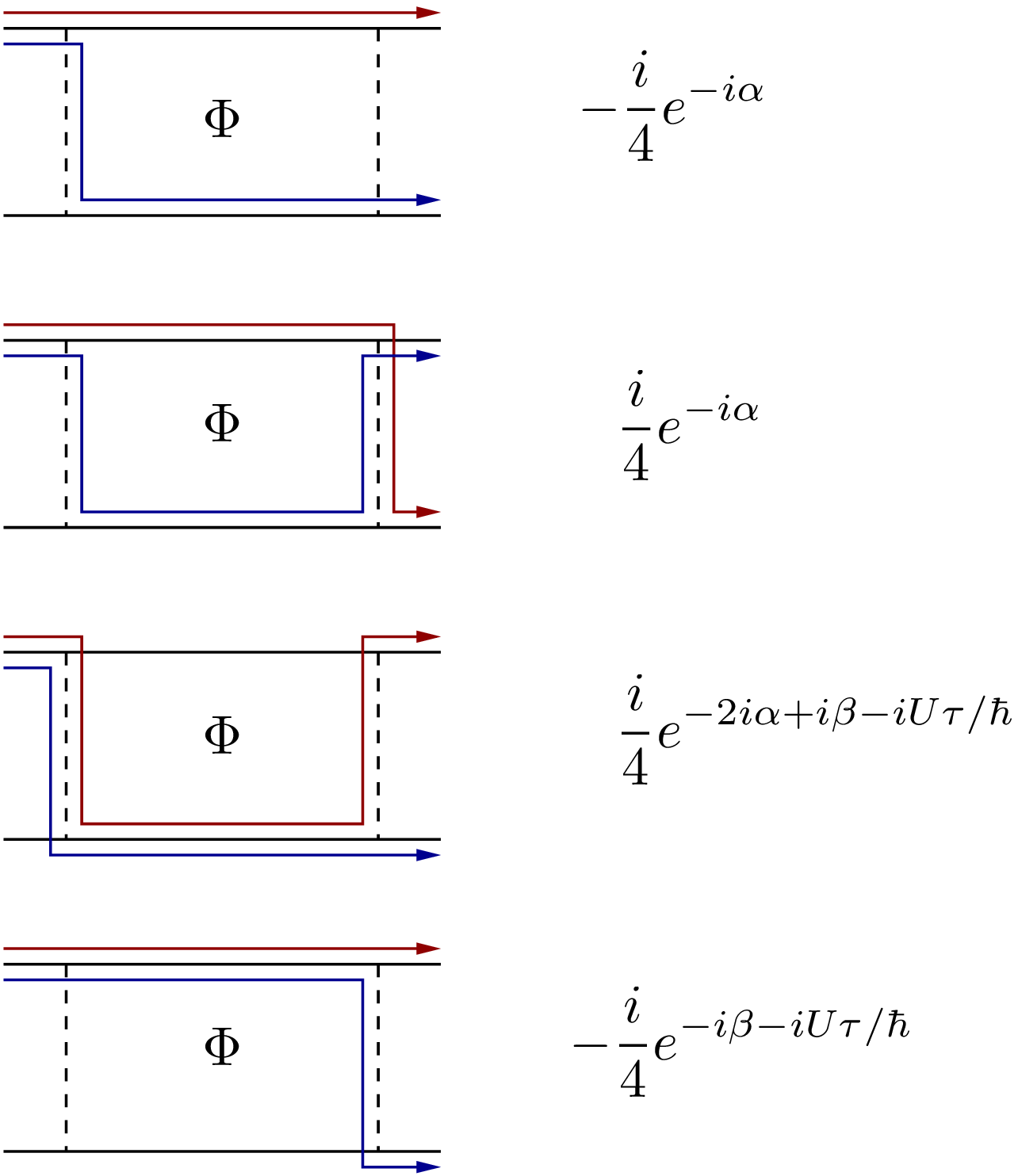,width=5.8cm} 
%\vspace*{-0.5cm}
\caption{(Color online) As in Fig. \protect\ref{fig_two_part_paths_1}, but
with one particle on the lower channel in the final state. The combined
amplitude is  $A_1=-\frac{1}{2}e^{-i\protect\alpha}e^{-iU\protect\tau%
/\hbar}\sin\Phi$.}
\label{fig_two_part_paths_2}
\end{figure}

Now consider the two-particle problem with an initial state as described, in which both particles are on the upper channel. Paths to a final state with both particles in the lower channel are shown in Fig.~\ref{fig_two_part_paths_1} and those to a final state with one particle in each channel in Fig.~\ref{fig_two_part_paths_2}. (Because there is no dispersion, particles cannot exchange positions in the scattering process.
They may therefore be treated as if they were distinguishable, and this is  reflected in
the figures by the use of different colours for the paths of each particle.)
Without interactions the amplitude for a given pair of paths would simply be a product of contributions for each particle.
Interactions contribute additional phase factors $e^{-iU \tau/\hbar}$ when both particles
propagate  on the same channel inside the interferometer.
The average charge transferred between the upper and lower channels in the scattering process is
\begin{equation}  \label{current_2part}
I = 2|A_2|^2+2|A_1|^2=1+\cos(U\tau/\hbar)\cos\Phi,
\end{equation}
where both terms are multiplied by factors of two, since $A_2$ describes two-particle transmission, while single-particle transmission with amplitude $A_1$ can occur for either particle. AB oscillations are represented in this expression by the term in $\cos\Phi$.
Their strength is modulated by the factor $\cos(U\tau/\hbar)$. We can take the interaction strength $U$ to play the same role in  the two-particle problem as bias voltage in the many-body system, since increasing bias leads to reduced spatial separation between electrons
entering the MZI above a filled Fermi sea, which in turn increases the interaction energy between these electrons.

The phenomenon can be summarised by defining the visibility of AB oscillations.  Let $I_{\rm max}$ and $I_{\rm min}$ be the maximum and minimum values of $I$ as $\Phi$ varies. The visibility is
\begin{equation}
{\cal V}=\frac{I_{\rm max}-I_{\rm min}}{I_{\rm max}+I_{\rm min}}.
\end{equation}
From Eq.~(\ref{current_2part}) we have ${\cal V} = |\cos(U\tau/\hbar)|$, and hence a lobe pattern in $\cal V$ as a function of $U$.
It is also evident from Eq.~(\ref{current_2part}) that the
phase of AB oscillations  changes abruptly by $\pi$ at 
zeros of $\cal V$.

We use the same approach to calculate the current for the case of
arbitrary tunneling amplitudes $t_a,t_b$ at the QPCs.
The result for the average charge transferred is
\begin{multline}
I=2[T_{a}R_{b}+R_{a}T_{b}+2(T_{a}T_{b}R_{a}R_{b})^{1/2}\times \\
\times \lbrack 1-4(T_{a}R_{a})\sin ^{2}(U\tau/\hbar)]^{1/2}\cos{\tilde\Phi%
}],
\end{multline}%
where $T_{a,b}=1-R_{a,b}=t^2_{a,b}$ is the tunneling probability.
The phase of AB oscillations is shifted  by interactions, being 
\begin{equation}
\tilde\Phi=\Phi+\arccos\{[1-4(T_{a}R_{a})\sin^{2}(U\tau/\hbar)]^{-1/2}\cos(U%
\tau/\hbar)\}.
\end{equation}%
The visibility is 
\begin{equation}
 {\cal V}={\cal V}_0\times\lbrack 1-4(T_{a}R_{a})\sin ^{2}(U\tau/\hbar)]^{1/2}
\end{equation}
where ${\cal V}_0$ is the single-particle value
\begin{equation}
 {\cal V}_0=\frac{2[T_a T_b R_a R_b]^{1/2}}{T_a R_b + R_a T_b}.
\end{equation}

An important difference between these results and the ones for the many body problem that we present in Section \ref{sec:application_potentials} is that here $\cal V$ does not decay at large $U$. In other respects, however, the two-body problem is illuminating. In particular, while the behaviour of $\cal V$ is not  affected by the value of $T_b$ except for a multiplicative factor, if $T_a \not= 1/2$  the zeros of the visibility turn into finite minima and the jumps in the phase of AB oscillations  become smooth rises. In the limit that transmission at the first QPQ approaches $T_a=1$ or $T_a=0$, modulations of $\cal V$ with $U$ disappear altogether. These features are also
present in the many body problem. One consequence is that the lobe pattern cannot be obtained at leading order from a calculation that is perturbative in tunneling.

\section{Microscopic model of the MZI}\label{sec:model}

\begin{figure}[tbp]
\epsfig{file=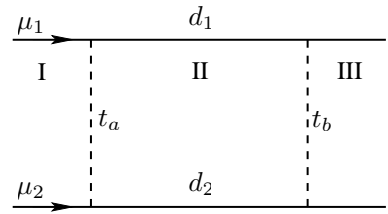,width=5cm} 
\caption{Schematic view of the
MZI. Horizontal lines represent edge
states with propagation direction indicated by arrows. These edge states are
connected by two QPCs, shown as vertical dashed lines, with
tunneling amplitudes $t_a$ and $t_b$. The arm lengths between contacts are
$d_{1}$ and $d_2$, and the chemical potentials in the incident channels are
$\protect\mu_{1}$ and $\mu_2$. The three different regions of the interferometer
discussed in the text are labeled using Roman numerals.}
\label{fig2}
\end{figure}

Our model for the MZI is sketched
in Fig. \ref{fig2}. The Hamiltonian is
\begin{equation} \label{ham}
\mathcal{\hat{H}}=\mathcal{\hat{H}}_{kin}+ \mathcal{\hat{H}}_{tun}+\mathcal{%
\hat{H}}_{int}\,.
\end{equation}
It has three contributions: $\mathcal{\hat{H}}_{kin}$ is the single particle term for an isolated edge; $\mathcal{\hat{H}}_{int}$ represents
electron-electron interactions; and $\mathcal{\hat{H}}_{tun}=\mathcal{\hat{H}}%
^{a}_{tun}+\mathcal{\hat{H}}^{b}_{tun}$ describes tunneling at the QPCs labelled 
$a$ and $b$.

We consider initially edge channels of length $L$ with periodic
boundary conditions, then take $L\to \infty$. Allowed wavevectors are $k=2\pi
n_{k}/L$, with $n_{k}$ integer. Fermionic operators $
\hat{c}_{k\eta}^{+}$ and $\hat{c}_{k\eta}$, which create and annihilate an electron with
momentum $k$ on the edge $\eta$, obey standard anticommutation
relations $\{\hat{c}_{k\eta},\hat{c}_{p\eta^{\prime}}^{+}\}=\delta_{kp}%
\delta_{\eta\eta^{\prime}}$. In coordinate representation the field operator $%
\hat{\psi}_{\eta}(x)$, which annihilates an electron at position $x$ on the
edge $\eta$, is 
\begin{equation}
\hat{\psi}_{\eta }(x)=\frac{1}{\sqrt{L}}\sum_{k=-\infty }^{\infty }\hat{c}%
_{k\eta }e^{ikx}\,. \label{psi}
\end{equation}

With this notation
\begin{equation}
\mathcal{\hat{H}}_{kin}=-i\hbar v_{F}\sum_{\eta =1,2}\int_{-L/2}^{L/2}\hat{%
\psi}_{\eta }^{+}(x)\partial _{x}\hat{\psi}_{\eta }(x)dx\,.  \label{H_kin}
\end{equation}%
We represent interactions within each edge using the same symmetric potential $U(x,x^\prime)$ and neglect interactions between electrons in different edges. Introducing
the density operator $\hat\rho_{\eta}(x)=\hat\psi^{+}_{\eta}(x)\hat\psi_{\eta}(x)$, we then have
\begin{equation}
\mathcal{\hat{H}}_{int}=\frac{1}{2}\sum_{\eta
=1,2}\int_{-L/2}^{L/2}U(x,x^{\prime })\hat{\rho}_{\eta }(x)\hat{\rho}_{\eta
}(x^{\prime })dxdx^{\prime }\,.  \label{H_int}
\end{equation}
Finally, taking the QPCs to be point-like, we write
\begin{eqnarray}
\mathcal{\hat{H}}^{a}_{tun}&=&v_{a}e^{i\alpha }\hat{\psi}_{1}^{+}(0)\hat{\psi}%
_{2}(0) +\mathrm{h.c.}, \label{H_tun_a}  \\
\mathcal{\hat{H}}^{b}_{tun}&=&v_{b}e^{i\beta }\hat{\psi}_{1}^{+}(d_{1})\hat{%
\psi}_{2}(d_{2})+\mathrm{h.c.}  \label{H_tun_b}
\end{eqnarray}
Here $v_{a}$ and $v_b$ are tunneling strengths, from which the quantum amplitudes $t_a$ and $t_b$ can be calculated. As in the previous section, the AB phase due to enclosed flux is
$\Phi=\beta-\alpha $.

This model can be solved exactly when interactions  occur only between pairs of 
electrons that are both inside the interferometer, in the region denoted II in Fig.~\ref{fig2}. We present results in Section \ref{sec:application_potentials} calculated using three different choices for such internal interactions. The first of these is simply a charging energy
\begin{equation}  \label{pot_range_ind}
U\left( x,x^{\prime }\right) =\left\{ 
\begin{array}{cc}
g & \quad 0<x,x^\prime<d \\ 
0 & \quad \mathsf{otherwise}\,.
\end{array}
\right.
\end{equation}
An interaction of this kind is standard in the theory of quantum dots and for an MZI
was treated approximately in Ref.~\onlinecite{neder08}. To test the robustness of behaviour to changes in the form of interaction, we also obtain results for two types of interaction potential that vary with electron separation inside the interferometer, taking
\begin{equation}  
U\left( x,x^{\prime }\right) =\left\{ 
\begin{array}{cc}
U(x-x^\prime) & \quad 0<x,x^\prime<d \\ 
0 & \quad  \mathsf{otherwise}%
\end{array}%
\right.
\end{equation}
and either an exponential dependence
\begin{equation}  \label{exp_inter}
U(x-x^{\prime})=g e^{-\alpha|x-x^{\prime}|}
\end{equation}
or a Coulomb form
\begin{equation}  \label{coulomb_int}
U(x-x^{\prime})=\frac{g_c}{\sqrt{(x-x^{\prime})^2+a_c^2}}\,.
\end{equation}

\section{Exact solution}\label{sec:exact-solution}

In this Section we give a full account of the solution outlined previously in
Ref.~\onlinecite{exact09}.
We study the interferometer at finite bias voltage by computing
the quantum-mechanical time evolution of an initial state in which 
the single-particle levels of $\mathcal{\hat{H}}_{kin}$ for the two edges
are occupied up to different chemical potentials, $\mu_1$ and $\mu_2$. 
The interferometer reaches a steady state at long times, and we evaluate observables in this state. 
The challenge of the calculation comes from the difficulty of treating both
tunneling and interactions non-perturbatively.
Each part of the problem can be described by a quadratic Hamiltonian, but the
appropriate variables are different in the two cases: fermionic for tunneling, and bosonic for interactions.  A model which has only internal interactions can be solved exactly 
because in these circumstances the effects of tunneling at
QPCs and of interactions can be handled separately. 
In the following we consider only edge states at filling factor $\nu{=}1$ and an initial state at zero temperature, but both these restrictions could be lifted within the approach.

To treat time evolution we use the interaction representation, taking as the free part of the Hamiltonian
\begin{equation}
\hat{H}_{0}=\mathcal{\hat{H}}_{kin}+\mathcal{\hat{H}}_{int}\,,
\end{equation}
and as the `interaction' part $\mathcal{\hat{H}}_{tun}$. In this representation the time evolution of the fermion operators is given by
\begin{equation}
\hat{\psi}_{\eta }(x,t)=e^{i\hat{H}_{0}t/\hbar }\hat{\psi}_{\eta }(x)e^{-i
\hat{H}_{0}t/\hbar }\,.
\end{equation}
The wavefunction of the system, which we denote at $t=0$ by $|\mathrm{Fs}\rangle $, 
evolves with the  S-matrix 
\begin{equation}  \label{Smat}
\hat{S}(t)=\mathrm{T}\exp \left\{ -\frac{i}{\hbar }\int_{0}^{t}\hat{\mathcal{H}}
_{tun}(\tau )d\tau \right\}\,,
\end{equation}
where $\mathrm{T}$ denotes time-ordering. We distinguish operators in the Schr\"{o}dinger and
interaction representations by the absence or presence of a time argument.
In Section \ref{sub:exact_cur_der} we also use operators in the Heisenberg representation, and we indicate these with a 
subscript $H$. 

The presentation of the remainder of the calculation is organised as follows.
The observable we are concerned with is the current through the MZI, and we derive a convenient form for the corresponding operator in Section \ref{sub:exact_cur_der}.  The simplifications arising in a model with only internal interactions affect the calculation of the $S$-matrix of Eq. (\ref{Smat}), which we describe in Section \ref{sub:exact_smat_eval}. 
To find the time evolution of fermion operators we use bosonization, as set out in Section \ref{bosonization}. After bosonization $\hat{H}_{0}$ is quadratic and may 
either be 
treated using scattering theory (Section \ref{scattering}) or
diagonalised using a Bogoliubov transformation (Section \ref{sub:exact_diag_ham}).
Inverting our transformations, we arrive in Section  \ref{sub:exact_eval_corr}
at an expression for current at long times, written in terms of fermion operators in the Schr\"odinger picture, and show that this expression is suitable for numerical evaluation. We give results for different choices of interaction potential in Section %\ref{sub:exact_eval_corr} and 
\ref{sec:application_potentials}.

\subsection{Derivation of the current operator}\label{sub:exact_cur_der}

Although it is usual to write $\mathcal{\hat{H}}_{tun}$ as in Eqns.~(\ref{H_tun_a}) and (\ref{H_tun_b}), this is a shorthand since at finite tunneling strength  the fermion field is discontinuous at QPCs. For that reason we regularise $\mathcal{\hat{H}}_{tun}$ by considering QPCs of finite extent $w$, taking $w\to 0$ at the end of calculations. Then, for example, $\mathcal{\hat{H}}_{tun}^{b}$ has the form
\begin{equation}
\mathcal{\hat{H}}_{tun}^{b}=\frac{1}{w}\int_{0}^{w}dx[v_{b}e^{i\beta }\hat{
\psi}_{1}^{+}(d_{1}+x)\hat{\psi}_{2}(d_{2}+x)+\mathrm{h.c.}]\,.
\end{equation}
The current operator can be found in the standard way from the number operator $\hat{N}_{1}=\int \hat{\rho}_{1}(x)dx$ for electrons on the upper edge, by evaluating its commutator with $\mathcal{\hat{H}}$. This gives for the current at QPC $b$ 
\begin{equation}  \label{current_b}
\hat{I}_{b}=-\frac{2e}{w\hbar}\Im\int_{0}^{w} v_{b}e^{i\beta }\hat{\psi}
_{1}^{+}(d_{1}+x)\hat{\psi}_{2}(d_{2}+x)dx
\end{equation}
and  a similar expression for $\hat{I}_{a}$, the current at QPC $a$. To calculate the position dependence of fermion fields within the QPCs we introduce 
operators in the Heisenberg representation, with the time dependence 
\begin{equation}
\hat{A}_H(t)=e^{i\mathcal{\hat{H}}t/\hbar}\hat A e^{-i\mathcal{\hat{H}}t/\hbar}\,.
\end{equation}
The equations of motion for the fermion operators at  the contact $b$ have
in the interval $0\leq x\leq w$ the form
\begin{align*}
(\partial _{t}+v_{F}\partial _{x})\hat{\psi}_{1H}^{+}(d_{1}+x,t)& =i\frac{%
v_{b}}{w}e^{-i\beta }\hat{\psi}_{2H}^{+}(d_{2}+x,t),\\
(\partial _{t}+v_{F}\partial _{x})\hat{\psi}_{2H}^{+}(d_{2}+x,t)& =i\frac{%
v_{b}}{w}e^{i\beta }\hat{\psi}_{1H}^{+}(d_{1}+x,t).  \notag
\end{align*}%
This system of equations has in the same interval the solution 
\begin{equation}  \label{psitrans_M}
\hat{\psi}_{\eta H}^{+}(d_{\eta }+x,t)=\sum_{\eta ^{\prime }=1,2}M_{\eta
\eta ^{\prime }}\hat{\psi}_{\eta ^{\prime }H}^{+}(d_{\eta ^{\prime }},t- x/v_F)\,,
\end{equation}
where $M=\exp (ix\Sigma )$ and the matrix $\Sigma $ is 
\begin{equation*}
\Sigma =\frac{v_{b}}{wv_{F}}\left( 
\begin{array}{cc}
0 & e^{-i\beta } \\ 
e^{i\beta } & 0%
\end{array}%
\right).
\end{equation*}%
An explicit expression for the matrix $M$ is  
\begin{equation*}
M=\left( 
\begin{array}{cc}
\cos \theta (x) & ie^{-i\beta }\sin \theta (x) \\ 
ie^{i\beta }\sin \theta (x) & \cos \theta (x)%
\end{array}%
\right),
\end{equation*}%
with $\theta (x)=xv_{b}/wv_{F}\hbar$.
Taking the limit $w\rightarrow 0$ and writing $\varepsilon =+0$, 
we arrive at a relation between
 fermion operators at $d_{\eta
}+\varepsilon $, just after the tunneling contact $b$, and those
at $d_{\eta ^{\prime }}-\varepsilon $, just before the contact:
\begin{equation}
\hat{\psi}_{\eta H}^{+}(d_{\eta }+\varepsilon ,t)=\sum_{\eta ^{\prime }=1,2}%
\mathcal{S}_{\eta \eta ^{\prime }}^{(b)\ast }\hat{\psi}_{\eta ^{\prime
}H}^{+}(d_{\eta ^{\prime }}-\varepsilon ,t),  \label{sc_transform}
\end{equation}%
with the scattering matrix
\begin{equation}
\mathcal{S}^{(b)}\equiv \left( 
\begin{array}{cc}
r_{b} & -it_{b}e^{i\beta } \\ 
-it_{b}e^{-i\beta } & r_{b}%
\end{array}%
\right) .  \label{s_matrix1}
\end{equation}%
In this way the reflection $r_{a,b}=\cos \theta _{a,b}$ and transmission $%
t_{a,b}=\sin \theta _{a,b}$ amplitudes at the contact $a,b$ are expressed in terms of the
angles $\theta _{a,b}=v_{a,b}/\hbar v_{F}$.

Substitution of Eq.~(\ref{psitrans_M}) into Eq.~(\ref
{current_b}) and reversion to the Schr\"odinger picture  gives for the current operator $\hat{I}_{b}$ the result
\begin{multline}
\hat{I}_{b }=ev_{F}\{t_{b}^{2}[\hat{\rho}_{1H}(d_{1}-\varepsilon )-%
\hat{\rho}_{2H}(d_{2}-\varepsilon )] \\
+t_{b}r_{b}[ie^{i\beta }\hat{\psi}_{1H}^{+}(d_{1}-\varepsilon )\hat{\psi}%
_{2H}(d_{2}-\varepsilon )+\mathrm{h.c.}]\},  \label{eq13}
\end{multline}%
in which we show explicitly that all operators are evaluated at
a point infinitesimally before the contact. In the
following we will omit $\varepsilon$. An expression for the
current operator at QPC $a$ is obtained from Eq.~(\ref{eq13}) by substituting $0$ for $d_{1,2}$ 
and replacing $\theta_b
$ with $\theta_a$.
The total current operator is then
$\hat{I}=\hat{I}_{a}+\hat{I}_{b}$. 

When considering expectation values, denoted by $\langle \ldots \rangle $ or by omitting hats, 
it is useful to separate the contribution from QPC $b$ 
into two terms, $I_{b}=I_{b}^{(1)}+I_{b}^{(2)}$, with 
\begin{eqnarray}  \label{IaIb}
I_{b}^{(1)} &=&ev_{F}t_{b}^{2}\langle \hat{\rho}_{1}(d_{1})-\hat{\rho}%
_{2}(d_{2})\rangle  \notag \\
I_{b}^{(2)} &=&ev_{F}t_{b}r_{b}[ie^{i\beta }\langle \hat{G}_{12}\rangle +%
\mathrm{h.c.}],
\end{eqnarray}%
where $\hat{G}_{12}=\hat{\psi}_{1}^{+}(d_{1})\hat{\psi}_{2}(d_{2})$.
The term $I_b^{(2)}$
is sensitive to the coherence between edges while $I_b^{(1)}$ is insensitive.
Since there is no coherence between channels before contact $a$, 
$I_{a}^{(2)}=0$ and
the contribution to the current from this contact is  
\begin{equation}  \label{IA}
I_{a}^{(1)}=ev_{F}t_{a}^{2}\langle \hat{\rho}_{1}(0)-\hat{\rho}%
_{2}(0)\rangle.
\end{equation}
The term responsible for AB-fringes in the current is $I_{b}^{(2)}$ and
our general task is to calculate $\langle \hat{G}_{12}\rangle$.

In experiment the differential conductance $\mathcal{G}=e\mathrm{\ d}I/%
\mathrm{d}\mu _{1}$ (with $\mu _{2}$ fixed) is measured at finite bias
voltage $V=(\mu _{1}-\mu _{2})/e$. $\mathcal{G}$ oscillates with $\Phi$, having maximum and minimum values $\mathcal{G}_{\max }$ and $%
\mathcal{G}_{\min }$. The \textit{AB fringe visibility} is defined as 
\begin{equation}
\mathcal{V}=\frac{\mathcal{G}_{\max }-\mathcal{G}_{\min }}{\mathcal{G}_{\max
}+\mathcal{G}_{\min }}.
\end{equation}

\subsection{Evaluation of the S-matrix}\label{sub:exact_smat_eval}

We require the action of the $S$-matrix, Eq.~(\ref{Smat}), on the initial
state $|Fs\rangle$. This state is represented by a product of fermion
creation operators acting on the vacuum and we need to find how $%
\hat{S}(t)$ transforms the fermion operators. Evaluation of $\hat{S}(t)$ is
based on our restriction of interactions to the interior of the MZI.
Specifically, separating $\mathcal{\hat{H}}_{tun}$ into parts $\mathcal{\hat{%
H}}_{tun}^{a}$ and $\mathcal{\hat{H}}_{tun}^{b}$ due to each QPC, we find
(see Appendix \ref{app:comm_rel}) that 
\begin{equation}
\lbrack \mathcal{\hat{H}}_{tun}^{a}(t_{1}),\mathcal{\hat{H}}%
_{tun}^{b}(t_{2})]=0
\end{equation}%
for $t_{1}\geq t_{2}$. This leads to a factorization of the
S-matrix into the product $\hat{S}(t)=\hat{S}^{b}(t)\hat{S}^{a}(t)$, where $%
\hat{S}^{a}(t)$ is the $S$-matrix calculated using $\hat{H}%
_{tun}^{a}$, and $\hat{S}^{b}(t)$ using $\hat{H}_{tun}^{b}$. A second
commutator (see again Appendix \ref{app:comm_rel}) 
\begin{equation*}
\lbrack \hat{G}_{12}(t_{1}),\mathcal{\hat{H}}_{tun}^{b}(t_{2})]=0,
\end{equation*}%
also valid for $t_{1}\geq t_{2}$, ensures that
\begin{equation}
[\hat{S}^{b}(t)]^{+}\hat{G}_{12}(t)\hat{S}^{b}(t)=\hat{G}_{12}(t),
\end{equation}
so an explicit form for $\hat{%
S}^{b}(t)$ is not required in the calculation. Since QPC $a$ acts before the
interacting region, it is easy to evaluate $\hat{S}^{a}(t)$ (Appendix \ref{app:s-matrix}):
we have 
\begin{equation*}
\lbrack \mathcal{\hat{H}}_{tun}^{a}(t_{1}),\mathcal{\hat{H}}%
_{tun}^{a}(t_{2})]=0
\end{equation*}%
for any $t_{1},t_{2}\geq 0$ and so may omit time ordering. The action of $%
\hat{S}^{a}(t)$ on fermionic operators is a rotation in the space of
channels and can be written as 
\begin{equation}
\tilde{\hat{\psi}}_{\eta }(x)=[\hat{S}^{a}(t)]^{+} \hat{\psi}_{\eta ^{\prime
}}(x)\hat{S}^{a}(t).
\end{equation}
For $0<x<v_{F}t$ we find the transformation 
\begin{equation}
\tilde{\hat{\psi}}_{\alpha }(x) =\sum_{\beta }\mathcal{S}_{\alpha \beta }^{a}%
\hat{\psi}_{\beta }(x),  \label{Sa}
\end{equation}
with the rotation matrix given by 
\begin{equation}  \label{scatt_mat_a}
\mathcal{S}^{a}=\left( 
\begin{array}{cc}
r_{a} & -it_{a}e^{i\alpha } \\ 
-it_{a}e^{-i\alpha } & r_{a}%
\end{array}%
\right).
\end{equation}

\subsection{Bosonization}
\label{bosonization}

To compute the time evolution of operators in the interaction
representation under $\hat{H}_{0}$ we use bosonization.\cite{vonDelft} This
gives us an exact correspondence between fermion and boson operators via
the bosonization identity 
\begin{equation}
\hat{\psi}_{\eta }(x)=(2\pi a)^{-1/2}\hat{F}_{\eta }e^{i\frac{2\pi }{L}\hat{N%
}_{\eta }x}e^{-i\hat{\phi}_{\eta }(x)},  \label{boson_id}
\end{equation}%
where bosonic fields are defined as 
\begin{equation}
\hat{\phi}_{\eta }\left( x\right) =-\sum_{q>0}\left( 2\pi /qL\right)
^{1/2}(e^{iqx}\hat{b}_{q\eta }+\mathrm{h.c.})e^{-qa/2} \label{phi_x}
\end{equation}%
and $a$ is an infinitesimal regulator, which does not enter the final
results. The plasmon creation and annihilation operators (which have $q >0$) obey bosonic commutation relations 
\begin{equation}
[\hat{b}_{q\eta },\hat{b}_{k\eta ^{\prime }}^{+}]=\delta _{qk}\delta _{\eta
\eta ^{\prime }}.
\end{equation}
They can be expressed in terms of fermions as 
\begin{equation}
\hat{b}_{q\eta }^{+}=i\left( 2\pi /qL\right) ^{1/2}\sum_{k=-\infty }^{\infty
}\hat{c}_{k+q\eta }^{+}\hat{c}_{k\eta }\;.  \label{b_op}
\end{equation}%
The commutation relations for the fields $\hat{\phi}_{\eta }\left( x\right) $ 
(omitting terms proportional to $1/L$: see
discussion in Ref.~\onlinecite{vonDelft}) read
\begin{equation*}
\lbrack \hat{\phi}_{\eta }(x),\partial _{y}\hat{\phi}_{\eta ^{\prime
}}\left( y\right) ]=-2\pi i\delta \left( x-y\right) \delta _{\eta \eta
^{\prime }}.
\end{equation*}
The Klein factors $\hat{F}_{\eta }$, which change fermion number by one,
satisfy the commutation relations 
\begin{equation*}
\{\hat{F}_{\eta },\hat{F}_{\eta ^{\prime }}^{+}\}=2\delta _{\eta \eta
^{\prime }},\ \lbrack \hat{N}_{\eta },\hat{F}_{\eta ^{\prime }}]=-\delta
_{\eta \eta ^{\prime }}\hat{F}_{\eta },
\end{equation*}%
with the standard expression for the particle number operator 
\begin{equation}  \label{number_op}
\hat{N}_{\eta}\equiv \sum_{k=-\infty}^{\infty}\hat{c}_{k\eta}^{+}\hat{c}%
_{k\eta}- \sum_{k=-\infty}^{\infty}\langle 0|\hat{c}_{k\eta}^{+}\hat{c}%
_{k\eta}|0\rangle
\end{equation}
in the edge $\eta$. Here the vacuum state $%
|0\rangle$ satisfies 
\begin{align}
\hat c_{k\eta}^{+}|0\rangle\equiv&0,\; k\le0 \nonumber\\
\hat c_{k\eta}|0\rangle\equiv&0,\; k>0 \nonumber \,.
\end{align}
The commutators
$[F_{\eta},\hat b_{k\eta^{\prime}}]$ and $\ [\hat N_{\eta},\hat
b_{k\eta^{\prime}}]$ are zero.
%\end{equation}
The electron density operator is given by
\begin{equation}  \label{density_bos}
\hat{\rho}_\eta\left( x\right) =-\frac{1}{2\pi }\partial _{x}\hat{\phi}%
_\eta\left( x\right) +\hat{N}_{\eta}/L.
\end{equation}

Since $\hat{H}_{0}$ does not couple channels, in the following we restrict
our attention to a single channel and omit channel labels until we reach
Section \ref{sub:exact_eval_corr}. The kinetic energy $\mathcal{\hat{H}}_{kin}$ for a single edge
in bosonized form is 
\begin{equation}
\mathcal{\hat{H}}_{kin}=\frac{\hbar v_{F}}{2}\int_{-L/2}^{L/2}\frac{dx}{2\pi 
}(\partial _{x}\hat{\phi}\left( x\right) )^{2}+\frac{2\pi }{L}\frac{\hbar
v_{F}}{2}\hat{N}(\hat{N}+1)\,.  \label{H_kin_boson}
\end{equation}%
Similarly, $\mathcal{\hat{H}}_{int}$ is quadratic, and given by
\begin{equation}
\mathcal{\hat{H}}_{int}=\frac{1}{2}\int_{0}^{d}\int_{0}^{d}U(x,x^{\prime })\hat{\rho}%
\left( x\right) \hat{\rho}\left( x^{\prime }\right) dxdx^{\prime }.
\label{H_int_boson}
\end{equation}%
\label{H0_boson}
Using this form of the Hamiltonian, our objective is to express the time-dependent boson field $\hat{\phi}(d,t)$ in the interaction representation, in terms the boson operators $\hat{b}_q$ and $\hat{b}^+_q$ in the Schr\"odinger representation. We set out two approaches to this calculation. One is based on the formalism of scattering theory. We use this to treat interactions for which we can obtain simple expressions for plasmon scattering phase shifts. The other is based on a Bogoliubov transformation. We use it to study Coulomb interactions.
 
\subsection{Scattering approach}
\label{scattering}

The theory of plasmon scattering in spatially inhomogeneous systems of quantum Hall edge channels has been studied quite extensively. An early treatment of a Hall bar is given in Ref.~\onlinecite{oreg} and a recent application to an MZI is described in Ref.~\onlinecite{sukhorukov07}. For the model we are concerned with, consider the equation of motion 
\begin{equation}
i\hbar \partial _{t}\hat{\phi}(x,t)=[\hat{\phi}(x,t),{\hat{H}}_{0}]\,.
\label{eqm_bos_phi}
\end{equation}
We separate 
%the field operator $\hat\phi(x,t)$ into two parts
$\hat{\phi}\left( x,t\right)= \hat{\phi}^{(0)}\left( x,t\right)+\hat{\phi}^{(1)}\left( x,t\right)$ into a part
$\hat{\phi}^{(0)}\left( x,t\right)$, proportional to $\hat{N}$, and another part  $\hat{\phi}^{(1)}\left( x,t\right)$,
independent of $\hat{N}$. They obey
\begin{eqnarray}
i \hbar \partial_t \hat{\phi}^{(0)}\left( x,t\right) &=& -i\hbar v_F \partial_x \hat{\phi}^{(0)}\left( x,t\right)  
+i\frac{\hat{N}}{L}\int_0^d U(x,y){\rm d}y
\nonumber\\
&& -\frac{i}{2\pi}\int_0^d U(x,y) \partial_{y} \hat{\phi}^{(0)}\left(y,t\right) {\rm d}y
 \end{eqnarray}
and
%$i\partial_t\hat\phi^{(1)}(x,t)=[\phi^{(1)}(x,t),\hat H_0]$, or in explicit form
\begin{eqnarray}
i \hbar \partial_t \hat{\phi}^{(1)}\left( x,t\right) &=& -i\hbar v_F \partial_x \hat{\phi}^{(1)}\left( x,t\right)
\nonumber\\
 &&  -\frac{i}{2\pi}\int_0^d U(x,y) \partial_{y} \hat{\phi}^{(1)}\left(y,t\right) {\rm d}y\,\,\,
 \label{e-o-m-phi1}
\end{eqnarray}
with initial conditions $\hat{\phi}^{(0)}\left( x,0\right)=0$ and $\hat{\phi}^{(1)}\left( x,0\right)=\hat{\phi}(x)$. Our aim is to find the Green function for Eq.~(\ref{e-o-m-phi1}).

The basis functions for a mode expansion of $\hat{\phi}^{(1)}\left( x,t\right)$ obey
the time-independent Schr\"odinger equation 
\begin{equation}\label{eom-f}
\omega_p f_p(x) = - i v_F \partial_x f_p(x) -\frac{i}{2\pi \hbar} \int_0^d U(x,y) \partial_y f_p(y)
\end{equation}
and satisfy the orthonormality relation 
\begin{equation}
\int_{-L/2}^{L/2}f_{p}(x)\partial _{x}f_{q}^{\ast }(x)dx=-2\pi i\delta _{pq}\,.
\label{norm_f}
\end{equation}
The Green function can therefore be written as
\begin{equation}
K(x,y;t) = \frac{i}{2\pi} \sum_p f_p(x) \partial_y f^*_p(y) e^{-i \omega_pt}
\end{equation}
and we have
\begin{equation}
\hat{\phi}^{(1)}\left( x,t\right) = \int_{-L/2}^{L/2}  K(x,y;t) \hat{\phi}\left(y\right) {\rm d}y\,.
\end{equation}

Interactions within the MZI generate a frequency-dependent phase shift $\delta_p$ for plasmons, and the form of $f_p(x)$ on either side of the interaction region is (neglecting a correction to the normalisation that vanishes as $d/L \to 0$) 
\begin{equation*}
f_{p}\left( x\right) =-\left( \frac{2\pi }{q L}\right)
^{1/2}\left\{ 
\begin{array}{ll}
e^{i qx } &\quad  x\leq 0 \\
e^{i (qx-\delta _{q})} & \quad  x\geq d\,.
\end{array}%
\right.
\end{equation*}
With periodic boundary conditions at finite $L$, the allowed values of $q$ are fixed by the condition $f_p(-L/2) = f_p(L/2)$, and these determine the frequencies 
$\omega_p = v_Fq$. 
At long times and for large $L$, the quantity we require, $K(d,y;t)$, can be expressed solely in terms of these phase shifts as
\begin{equation}
K(d,y;t) = \frac{1}{2\pi} \int _{-\infty}^{\infty}{\rm d} p\, e^{i( p[d-y-v_Ft]-\delta_p)} \,.
\end{equation}
From this we obtain 
\begin{equation}
\hat{\phi}^{(1)}(x,t) = \sum_{q>0}(z_q(x,t) \hat{b}_q + {\rm h. c.})\,,
\end{equation}
in which the coefficients at long times have the form
\begin{equation}
z_{q}(d,t)=-\left( 2\pi /qL\right) ^{1/2}e^{iq(d-v_{F}t)-i\delta _{q}}\,.
\label{zq}
\end{equation}
The long time limit of $\hat{\phi}^{(0)}(d,t)$, which we write as $\hat{\phi}_0(d)$,
can also be expressed in terms of the phase shifts, as
\begin{equation}
\hat{\phi}_0(d) = 2 \pi \frac{\hat{N}}{L} \lim_{q\to 0} \frac{\delta_q}{q}\,.\label{phi0}
\end{equation}

\subsection{Diagonalisation by Bogoliubov transformation}\label{sub:exact_diag_ham}

An alternative approach is to diagonalise the Hamiltonian.
Substituting Eq. (\ref{phi_x}) into $\hat{H}_{0}$ we obtain 
\begin{multline}  \label{H0_bos}
\hat{H}_{0}=\hbar v_{F}\sum_{q>0}q(\hat{b}_{q}^{+}\hat{b}_{q}+1/2) \\
+\frac{\hbar }{2L}\sum_{k,q>0}\sqrt{qk}[u_{-q,-k}\hat{b}_{q}\hat{b}%
_{k}^{+}+u_{q,k}\hat{b}_{q}^{+}\hat{b}_{k} \\
-u_{q,-k}\hat{b}_{q}^{+}\hat{b}_{k}^{+}-u_{-q,k}\hat{b}_{q}\hat{b}_{k}]+\hat{%
H}_{N\phi }+\hat{H}_{N}
\end{multline}%
where the matrix elements of the interaction potential are 
\begin{equation*}
u_{q,k} =\frac{1}{2\pi \hbar }\int_{0}^{d}\int_{0}^{d}e^{-iqx+iky}U\left(
x,y\right) dxdy\,.
\end{equation*}
They obey the relations $u_{q,k}^{\ast }=u_{-q,-k}$ and $u_{q,k}=u_{-k,-q}$.

The last two terms in Eq.~(\ref{H0_bos}) involve the number operator $\hat N$. 
The first of them appears because of our choice of nonuniform interactions, which leads to a coupling 
\begin{equation}
\hat{H}_{N\phi }=i\hbar (\hat{N}/L)\left( 2\pi /L\right) ^{1/2}\sum_{k>0}
\sqrt{k}[u_{0,k}\hat{b}_{k}-u_{0,-k}\hat{b}_{k}^{+}]
\end{equation}
between the
plasmon and the number operators: by contrast,
in a system with translationally invariant interactions
there would be no such coupling. The  other term, $\hat{H}_{N}$, has the form
\begin{equation}  \label{number_hams}
\hat{H}_{N}=\frac{2\pi }{L}\frac{\hbar v_{F}}{2}\hat{N}(\hat{N}+1)+\frac{%
\hat{N}^{2}}{2L^{2}}\int_{0}^{d}U\left( x,x^{\prime }\right)
dxdx^{\prime }.
\end{equation}%
Since interactions in our model are limited to the finite region of length 
$d$, the second term in Eq.~(\ref{number_hams}) 
gives a correction to, for example, the equation of motion of the Klein factor
that is small in $d/L$ and so vanishes in the thermodynamic limit.
We therefore omit it in the following.

Contributions to the Hamiltonian linear in $\hat N$ are removed by making the shifts 
\begin{equation}
\hat{b}_{q}=\tilde{b}_{q}+
%(2\pi/qL)^{1/2}
{\alpha }_{q}\hat{N}\,,
\label{b_shifts}
\end{equation}
with coefficients $\alpha _{q}$ given by 
\begin{equation}
v_{F}\alpha _{q}+\frac{1}{L}\sum_{k>0}\sqrt{k/q}[u_{qk}\alpha
_{k}-u_{q,-k}\alpha _{k}^{\ast }]=\frac{i}{L}\left(\frac{2\pi}{qL}\right)^{1/2}\!\!\!u_{q0}\,. \label{eqalpha}
\end{equation}

The Hamiltonian, Eq.~(\ref{H0_bos}), written in terms of these shifted operators, is
diagonalised using a Bogoliubov
transformation of the form 
\begin{equation}
\hat{\beta}_{p}^{+}=\sum_{q>0}(A_{pq}\tilde{b}_{q}^{+}+B_{pq}%
\tilde{b}_{q}).  \label{b_trans1}
\end{equation}%
To preserve the commutation relations we require%
\begin{align}
\sum_{k>0}(A_{pk} A^{+}_{kq}-B_{pk}B^{+}_{kq})&= \delta_{pq} \\
\sum_{k>0}(B_{pk}A^{T}_{kq}-A_{pk}B^{T}_{kq})&=0,  \label{btrcond}
\end{align}%
which can be written in the matrix form%
\begin{equation}
\left( 
\begin{array}{cc}
A & B \\ 
B^{\ast } & A^{\ast }%
\end{array}%
\right) \left( 
\begin{array}{cc}
A^{+} & -B^{T} \\ 
-B^{+} & A^{T}%
\end{array}%
\right) =\left( 
\begin{array}{cc}
I & 0 \\ 
0 & I%
\end{array}%
\right),  \label{bog_mat}
\end{equation}%
where $I$ is the identity matrix.

The result is
\begin{equation}
\mathcal{H}_{b}=\sum_{p>0}\hbar \omega _{p}(\hat{\beta}_{p}^{+}\hat{\beta}%
_{p}+1/2)+\hat{H}_{N}+\mathrm{const.}  \label{ham_bos_b}
\end{equation}%
The time dependence of the transformed boson operators is given in the usual
way in terms of their frequencies $\omega _{p}$ as $\hat{\beta}%
_{p}(t)=e^{-i\omega _{p}t}\hat{\beta}_{p}$. Expressions for the
coefficients $A_{pq},B_{pq}$ can be found from the commutator 
\begin{equation}
\partial _{t}\hat{\beta}_{q}=-i[\hat{\beta}_{q},\hat{\mathcal{H}}%
_{b}]=-i\hbar \omega _{q}\hat{\beta}_{q}  \label{comm_beta}
\end{equation}%
which leads to the linear system of Bogoliubov equations%
\begin{eqnarray}
(\omega _{p}-v_{F}q)A_{pq} &=&\sum_{k>0}\sqrt{qk}%
[u_{qk}A_{pk}+u_{q,-k}B_{pk}]  \label{eq1AB} \\
(\omega _{p}+v_{F}q)B_{pq} &=&-\sum_{k>0}\sqrt{qk}%
[u_{-q,k}A_{pk}+u_{-q,-k}B_{pk}] \,. \notag
\end{eqnarray}

From Eq. (\ref{bog_mat}) we can obtain the inverse of the
Bogoliubov transformation, which we write in the interaction representation as%
\begin{equation}
\tilde{b}_{q}^{+}(t)=\sum_{p>0}(\hat{\beta}_{p}^{+}e^{i\omega
_{p}t}A_{pq}^{\ast }-\hat{\beta}_{p}e^{-i\omega _{p}t}B_{pq}).  \label{bbeta}
\end{equation}%
Substituting (\ref{b_trans1}) and (\ref{b_shifts}) into equation (\ref%
{bbeta}) we obtain the time dependence of the bosonic fields in terms of the
original operators $\hat{b}_{q}$ written in the Schr\"odinger representation
as 
\begin{equation}
\hat{\phi}\left( x,t\right) =\hat{\phi}^{(0)}\left( x,t\right)
+\sum_{q>0}(z_{q}(x,t)\hat{b}_{q}+\mathrm{h.c.}),  \label{phi_boson1}
\end{equation}%
where 
\begin{equation}  \label{zq0}
z_{q}(x,t)=\sum_{p>0}(A_{pq}^{\ast }f_{p}\left( x\right)
e^{-i\omega _{p}t}+B_{pq}f_{p}^{\ast }\left( x\right) e^{i\omega _{p}t})
\end{equation}%
with 
\begin{equation}  \label{fp}
f_{p}\left( x\right) =-\sum_{q>0}\left( 2\pi /qL\right)
^{1/2}(A_{pq}e^{iqx}-B_{pq}e^{-iqx}).
\end{equation}%
%We will later identify the functions $f_p(x)$ as
%coefficients in the mode expansion of the bosonic fields.
The term $\hat{\phi}^{(0)}\left( x,t\right) $ in Eq.~(\ref{phi_boson1}), arising from the operator
shifts, is given by 
\begin{equation}
\hat{\phi}^{(0)}\left( x,t\right) =\hat{\phi}_{0}\left( x\right) -\hat{N}%
\sum_{q>0}
%(2\pi/qL)^{1/2}
(z_{q}\left( x,t\right) \alpha _{q}+\mathrm{c.c.})
\label{eqphin}
\end{equation}%
with 
\begin{equation*}
\hat{\phi}_{0}(x)=-\hat{N}\sum_{q>0}(2\pi /qL)^{1/2}(\alpha _{q}e^{iqx}+%
\mathrm{c.c.}).
\end{equation*}%
It is easy to check that at $t=0$ the field (\ref{phi_boson1}) is equal to
the Schr\"{o}dinger operator $\hat{\phi}(x)$ of Eq.~(\ref{boson_id}
), since
%\begin{equation*}
$z_{q}(x,0)=-\left( 2\pi /q L\right) ^{1/2}e^{iqx} %,\text{ }\hat{\phi}%_{N}\left( x,0\right) =0.
$%\end{equation*}
 and $\hat{\phi}^{(0)}\left( x,0\right) =0$.
 
To make use of these results, the Bogoliubov coefficients $A_{pq}$, $B_{pq}$, 
$\alpha_q$ and frequencies $\omega _{p}$ are required. They can
be found from Eqns.~(\ref{eq1AB}) and (\ref{eqalpha}), using a numerical treatment with a momentum cutoff.

\subsection{Evaluation of the correlators}\label{sub:exact_eval_corr}

In this section we explain how the treatment we have described of the bosonized Hamiltonian enables evaluation of the correlator $\hat G_{12}(t)$, which appears
in Eq.~(\ref{IaIb}) and determines visibility of AB oscillations. 
When written using fermion fields  in the interaction representation, it is
\begin{equation}
\hat G_{12}(t)=\langle Fs|[\hat
S^{a}(t)]^{+}\hat\psi_1^{+}(d_1,t)\hat\psi_2(d_2,t)\hat S^{a}(t)|Fs\rangle\,. \label{G12}
\end{equation}
Our objective is to express $\hat\psi_1^{+}(d_1,t)$ and $\hat\psi_2(d_2,t)$ in terms 
fermion fields  $\hat{\psi}_{\eta }(x)$ and $\hat{\psi}_{\eta
}^{+}(x)$ in the Schr\"{o}dinger representation, so that the expectation value in the state $|Fs\rangle$ can be computed. To this end we use the bosonization identity, Eq.~(\ref{boson_id}), and simplify notation by defining
\begin{equation}
\hat{\mathcal{F}}_{\eta}(t)\equiv (2\pi a)^{-1/2}e^{i\hat H_0 t} \hat
F_{\eta} e^{i\frac{2\pi}{L}\hat N_\eta d_\eta}e^{-i\hat H_0 t}\,.
\end{equation}
Then 
\begin{equation}  \label{psi_F_phi}
\hat\psi_\eta(d_\eta,t)=\hat{\mathcal{F}}_{\eta}(t)e^{-i\hat\phi_{\eta}(d_%
\eta,t)}\,.
\end{equation}
The bosonic field $\phi_\eta(d_\eta,t)$ which appears here in the interaction representation is related at long times to one in the Schr\"odinger representation by Eqns.~(\ref{zq}) and (\ref{phi0}).
In addition, the time evolution of the operator $\hat{\mathcal{F}}_\eta(t)$ can be found in the usual way, via its commutator with $\hat H_0$
(the contribution from $\hat H_{N\phi}$ is small 
in $d/L$ and can be omitted in the thermodynamic limit), giving
\begin{equation}
\hat{\mathcal{F}}_{\eta}(t)=(2\pi a)^{-1/2}\hat F_{\eta} e^{i\frac{2\pi}{L}%
\hat N_\eta (d_\eta-v_F t)}.
\end{equation}
We wish to substitute for the Klein factor $\hat F_{\eta}$ in this expression. 
Consider a fermionic operator $\hat\psi_\eta(z_\eta)$ with $%
z_\eta=d_\eta-v_F t$. In the bosonized form, from Eq.~(\ref%
{boson_id}) it is
\begin{equation}
\hat{\psi}_{\eta }(z_\eta)=(2\pi a)^{-1/2}\hat{F}_{\eta }e^{i\frac{2\pi }{L}%
\hat{N}_{\eta }(d_\eta-v_F t)}e^{-i\hat{\phi}_{\eta }(z_\eta)}\,.
\label{boson_z_eta}
\end{equation}%
Multiplying Eq.~(\ref{boson_z_eta}) by $e^{i\hat{\phi}%
_{\eta }(z_\eta)}$ from the right we obtain ${\mathcal{F}}_\eta(t)$ in terms
of the fermion operators and bosonic fields in their Schr\"odinger
representation, as
\begin{equation}  \label{F_psi}
\hat{\mathcal{F}}_\eta(t)=\hat{\psi}_{\eta }(z_\eta)e^{i\hat{\phi}_{\eta
}(z_\eta)}=e^{i\hat{\phi}_{\eta }(z_\eta)}\hat{\psi}_{\eta }(z_\eta),
\end{equation}%
where the second equality holds due to the commutation relations of
Klein factors with bosonic operators.
Substituting Eq. (\ref{F_psi}) into Eq. (\ref{psi_F_phi}) we obtain 
\begin{equation}
\hat\psi_{\eta}(d_\eta,t)=e^{-i\hat\phi_{\eta}(d_\eta,t)}e^{i\hat\phi_{%
\eta}(z_\eta)}\hat\psi_{\eta}(z_{\eta})
\end{equation}
which can be written as 
\begin{equation}  \label{sol_psi}
\hat\psi_{\eta}(d_\eta,t)=e^{-i\varphi_\eta}e^{-i[\hat\phi_{\eta}(d_\eta,t)-%
\hat\phi_{\eta}(z_\eta)]}\hat\psi_{\eta}(z_{\eta})
\end{equation}
where the constant phase shift $\varphi_\eta$ is given by 
\begin{equation}
\varphi_\eta=\frac{i}{2}\int \frac{{\rm d}q}{q}e^{-i\delta_{q\eta}}\,.
\end{equation}
Finally, we substitute for $\hat{b}_{q\eta}$ and $\hat{b}_{q\eta }^{+}$ in $%
\hat{\phi}_{\eta }(x)$ in terms of fermion operators, using Eq.~(\ref{b_op}), with the result
\begin{equation}
\hat{\psi}_{\eta }(d_{\eta },t)=e^{-i\varphi_\eta}e^{-i\hat{Q}_{\eta }}\hat{%
\psi}_{\eta }(z_{\eta })\,.  \label{psi_dt}
\end{equation}%
Here the phase operator $\hat{Q}_{\eta }$ is 
\begin{equation}
\hat{Q}_{\eta }=\int_{-\infty }^{\infty }Q_{\eta }(x-z_{\eta })\hat{%
\rho}_{\eta }(x)dx,\label{phase-operator}
\end{equation}
where kernel $Q_{\eta}(x)=
\int {\rm d}q\,
\tilde{Q}_{\eta }(q)e^{iqx}$ has Fourier transform
\begin{equation}  \label{kernel_mom}
\tilde{Q}_{\eta}(q)=-\frac{i}{q}(e^{i\delta_{q\eta}}-1)\,.
\end{equation}

\begin{figure}[htp]
\epsfig{file=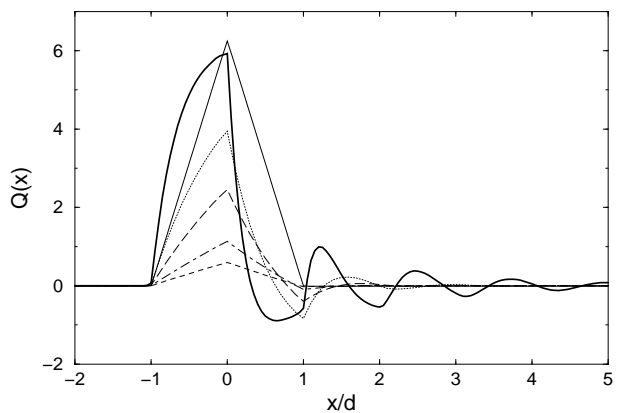,width=5.5cm,angle=270}
\caption{The kernel Q(x) of Eq. (\protect\ref{phase-operator}) for different
values of the interaction strength: $\protect\gamma=0.1$ (short-dashed
line), $\protect\gamma=0.2$ (dot-dashed line), $\protect\gamma=0.5$
(long-dashed line), $\protect\gamma=1.0$ (dotted line) and $\protect\gamma%
=3.0$ (full line) compared to that of Eq. (11) of Ref. \protect\onlinecite{neder08} at $\protect%
\gamma=1.0$ (thin full line)}
\label{fig_kernel}
\end{figure}

Eq.~(\ref{psi_dt})  is a key result which has a direct physical interpretation.
An electron passing through the interferometer accumulates a phase due to interactions with
other electrons. This phase is a collective effect and it is represented at the point where the electron leaves the MZI by
the operator $\hat{Q}_\eta$ in Eq.~(\ref{psi_dt}).
Contributions to the phase from the interactions with particles at a position $x$
from the QPC $b$ have a weight determined by the kernel $Q_\eta(x)$. The form of the kernel is illustrated 
in Fig.~\ref{fig_kernel} for the case of a charging interaction, studied in Section
~\ref{sub:range-independent}. The kernel has a maximum near $x=0$, showing that
interactions with nearby electrons are most important, but the phase is influenced
by all the electrons which have passed the interferometer, although with contributions which
decay with the distance $x$. The precise form of the kernel depends on the nature of the
interaction potential and reflects the full many-body physics of the problem.
A similar kernel appears in Eq.~(11) of Ref.~\onlinecite{neder08}, but with a simpler form because
of the approximations employed there. It is shown for comparison in Fig.~\ref{fig_kernel};
see discussion in Section~\ref{sec:discussion_neder}.

Substituting Eq.~(\ref{psi_dt}) into Eq. (\ref{G12}) we arrive at 
\begin{equation}
\langle \hat G_{12}(t)\rangle =e^{i\bar{\Phi}}\langle Fs|[\hat{S}^{a}(t)]^{+}%
\hat{\psi}_{1}^{+}(z_{1})e^{i\hat{R}}\hat{\psi}_{2}(z_{2})\hat{S}%
^{a}(t)\left\vert Fs\right\rangle.  \notag
\end{equation}%
Here $\hat{R}=\hat{Q}_{1}-\hat{Q}_{2}$ and $\bar{\Phi}=\varphi_1-\varphi_2$.
The action of $\hat{S}^{a}(t)^{+}$ and $\hat{S}^{a}(t)$ on the
operators they enclose is given by Eq. (\ref{Sa}). After this transformation
the correlator reads 
\begin{multline}\label{corr1}
\langle \hat G_{12}(t)\rangle=e^{i\bar{\Phi}}\frac{1}{L}\sum_{k,q=-\infty}^{%
\infty} {\mathcal{S}}_{1\alpha}^{a*}{\mathcal{S}}_{2\beta}^{a}e^{-ikz_1+i q
z_2}\times \\
\langle Fs|[\hat{c}_{k\alpha}^{+}e^{i\hat{\mathcal{R}}}\hat{c}_{q\beta}
\left\vert Fs\right\rangle,
\end{multline}
with summation over repeated indices $\alpha,\beta$. Here $\hat{%
\mathcal{R}}=[\hat{S}^{a}(t)]^{+}\hat{R}\hat{S}^{a}(t)$ is the rotated
kernel 
\begin{multline}\label{matR}
\hat{\mathcal{R}}={\mathcal{S}}_{1\alpha}^{a*}{\mathcal{S}}%
_{1\beta}^{a}\int_{-\infty}^{\infty} {\rm d}x\,Q_1(x-z_1)
\hat\psi_{\alpha}^{+}(z_1)\hat\psi_{\beta}(z_1) \\
-{\mathcal{S}}_{2\alpha}^{a*}{\mathcal{S}}_{2\beta}^{a}\int_{-\infty}^{%
\infty} {\rm d}x\, Q_2(x-z_2)\hat\psi_{\alpha}^{+}(z_2)\hat\psi_{\beta}(z_2).
\end{multline}
Now evaluation of $\langle \hat{G}_{12}(t)\rangle $ reduces to the
calculation of correlators of the form 
\begin{equation}\label{cmunu}
C_{\mu \eta }=\langle Fs|\hat{c}_{\mu }^{+}\exp ({i\sum_{\alpha \beta }%
\mathrm{M}_{\alpha \beta }\hat{c}_{\alpha }^{+}\hat{c}_{\beta }})c_{\eta
}|Fs\rangle ,
\end{equation}%
where the indices specify both channel and momentum, and the matrix $\mathrm{%
M}$ is obtained from $\hat{\mathcal{R}}$. One can show (see Appendix \ref{app:mat_els})
that $C_{\mu \eta }=\mathrm{D}_{\eta \mu }^{-1}\det \mathrm{D}$ with $%
\mathrm{D}$ constructed from the matrix elements of $\exp (i\mathrm{M})$
between the single-particle states that are occupied in the Slater
determinant $|Fs\rangle $. We calculate $C_{\mu \eta }$ numerically,
achieving convergence of the results when keeping up to $10^{3}$ basis
states and $400$ particles in each channel: further details are given in Appendix \ref{app:num_cor}.

\section{Results for various interaction potentials}\label{sec:application_potentials}

In this Section we apply our theory to study
interferometers with the three types of interaction potential introduced in Section \ref{sec:model}. 
In the absence of interactions the interferometer at finite
bias is specified by four dimensionless parameters: the tunneling probabilities $t_a^2$ and $t_b^2$
at the two QPCs;  the dimensionless bias $e V \sqrt{d_1d_2}
/2\pi\hbar v_F$; and the ratio of the arm lengths $d_2/d_1$. The tunneling probability $t_b^2$ at
the second contact QPC affects only the overall scale for visibility of AB oscillations, and
we set it to $t_b^2=1/2$. Interactions in general
introduce another parameter, characterising
their strength. Exponential and 
Coulomb interactions also depend on a further parameter: the interaction range
or the short-distance cutoff, respectively.

\subsection{Charging interaction}\label{sub:range-independent}\label{sub:exact_gen_const_int}

Consider first the charging interaction, Eq.~(\ref{pot_range_ind}).
It is characterized by the single dimensionless coupling constant $\gamma=gd/2\pi \hbar v_{F}$.
Solving Eq.~(\ref{eom-f}) we find 
\begin{equation*}
f_{p}\left( x\right) =-\left( \frac{2\pi }{qL}\right)
^{1/2}\left\{ 
\begin{array}{ll}
e^{iq x} & x\leq 0 \\ 
r_{p}+s_{p}e^{iqx} & 0<x<d \\ 
e^{iqx-i\delta _{q}} & x\geq d%
\end{array}%
\right. 
\end{equation*}%
with $q=\omega_p/v_F$.
Matching $f_{p}(x)$ at $x=0$ and $x=d$ gives $s_{p}=(1+t_{p})^{-1}$ and $r_{p}=t_{p}s_{p}$, with $%
t_{p}=(g/2\pi i\hbar \omega _{p})(e^{i\omega _{p}d/v_{F}}-1)$.
The phase shift $\delta
_{p}$ of plasmons due to the interactions is
\begin{equation}
e^{-i\delta _{p}}=(1+t_{p}^{\ast })/(1+t_{p}).  \label{phase_shifts}
\end{equation}%
Similarly, we find 
\begin{equation}
\hat{\phi}_{0}\left( x\right) =2\pi \bar{\gamma}\hat{N}x/L  \label{phi_zero}
\end{equation}%
for $0\leq x \leq d$, where $\bar{\gamma}=\gamma (1+\gamma )^{-1}$.
The contribution $-\frac{1}{2\pi }\partial _{x}\hat{\phi}_{0}\left( x\right) =-\bar{\gamma}\hat{N}/L$ to
the density inside the interferometer represents charge expulsion due to
interactions: in the limit of strong interactions the average density inside
the interferometer in the stationary regime is pinned at zero, independently
of $\hat{N}$. 

\begin{figure}[htb]
\epsfig{file=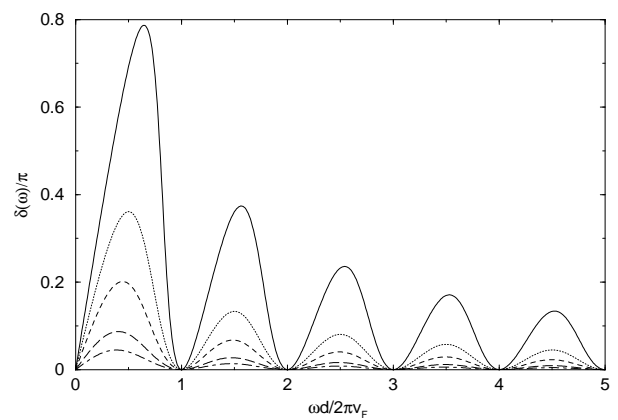,width=5.5cm,angle=270}
\caption{Frequency dependence of the plasmon phase shift for a
charging interaction with strength $\protect\gamma=0.1$
(dot-dashed line), $\protect\gamma=0.2$ (long-dashed line), $\protect\gamma%
=0.5$ (short-dashed line), $\protect\gamma=1.0$ (dotted line), and $\protect\gamma%
=3.0$ (full line). }
\label{fig_phase_shifts_const}
\end{figure}

The plasmon phase shift, as shown in Fig.~\ref{fig_phase_shifts_const}, varies linearly with frequency at low frequency and
falls to zero at high frequency. The maximum occurs at a frequency that increases with interaction strength and (for general interactions) depends on the shorter of two lengths: the interaction range and the arm length.
For strong interactions, the phase shift at fixed frequency approaches the limiting value $\delta_p = \omega_pd/v_F$. It then exactly cancels the kinetic phase $qd$.
This remarkable cancellation together with the charge expulsion results in behaviour 
independent of arm length when interactions are strong. A similar cancellation was found for a different model in Ref.~\onlinecite{sukhorukov08}.

\begin{figure}[htb]
\epsfig{file=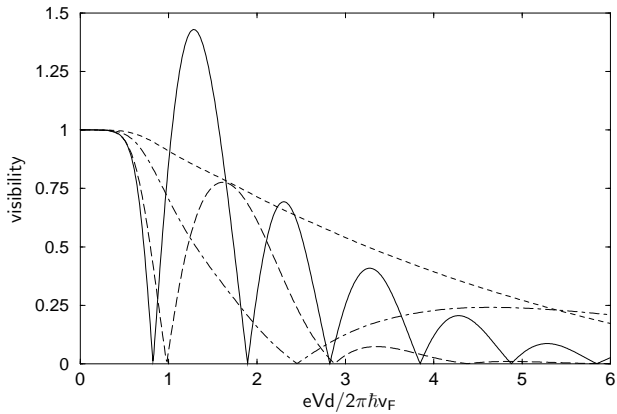,width=8cm,angle=0}
\caption{Visibility as a function of bias voltage for an MZI with 
charging interactions, and with $d_{1}=d_{2}$ and $%
t_{a}^{2}=t_{b}^{2}=1/2$, at interaction strengths: $\protect\gamma=0.1$
(short-dashed line), $\protect\gamma=0.2$ (dot-dashed line), $\protect\gamma%
=0.5$ (long-dashed line) and $\protect\gamma=1.0$ (full line). The phase of
the AB-fringes (not shown) jumps by $\protect\pi$ at zeros of the visibility.}
\label{fig_visibility_range_ind}%
\end{figure}
\begin{figure}[htb]
\epsfig{file=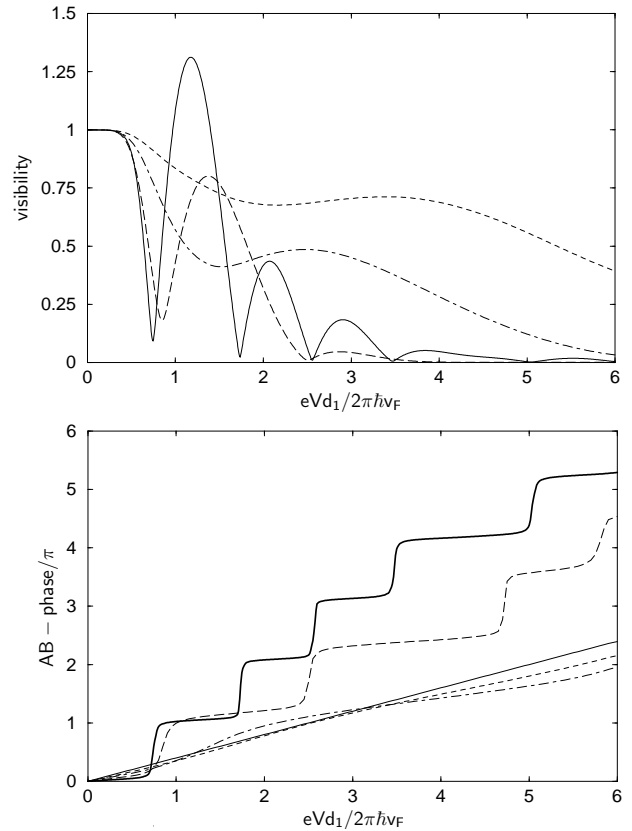,width=8cm,angle=0}
\caption{Visibility (upper panel) and AB phase (lower panel) as a
function of bias voltage for an MZI with $d_{2}/d_{1}=1.2$ and $%
t_{a}^{2}=t_{b}^{2}=1/2$ at interaction strengths: $\protect\gamma=0.1$
(short-dashed line), $\protect\gamma=0.2$ (dot-dashed line), $\protect\gamma%
=0.5$ (long-dashed line) and $\protect\gamma=1.0$ (full line). The linear
dependence of the AB-phase in the noninteracting case $\protect\gamma=0$ is
shown on the lower panel (thin full line).}
\label{fig_vis_nonequal_arms}%
\end{figure}
\begin{figure}[tbp]
\epsfig{file=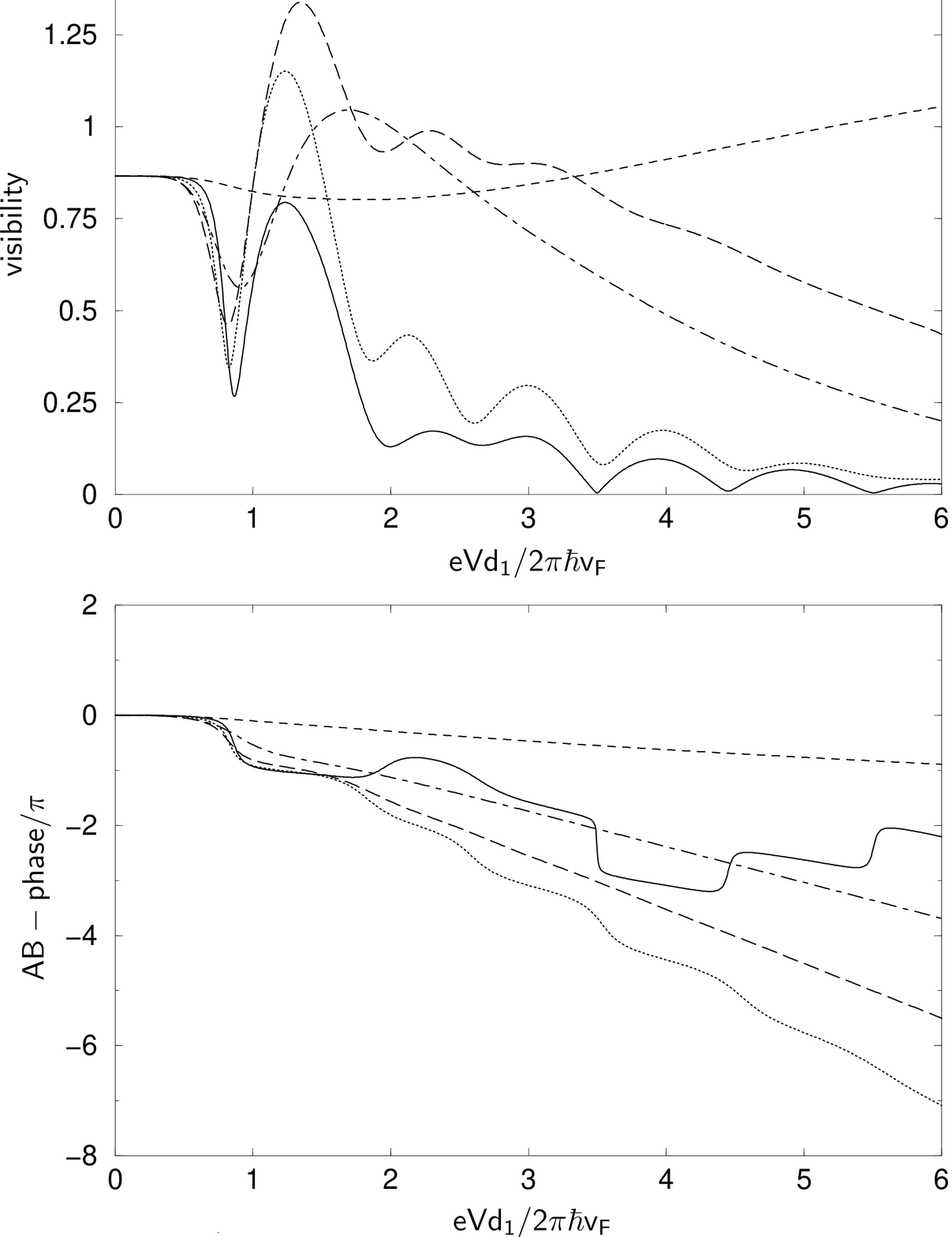,width=8cm,angle=0}
\caption{Visibility (upper panel) and the AB-phase (lower panel) as a
function of bias voltage for an MZI with $d_{2}=d_{1}$, $t_{a}^{2}=0.75$ and $%
t_{b}^{2}=1/2$, and charging interactions of
strength: $\protect\gamma=0.1$ (short-dashed line), $\protect\gamma=0.5$
(dot-dashed line), $\protect\gamma=1.0$ (long-dashed line), $\protect\gamma%
=2.0$ (dotted line) and $\protect\gamma=3.0$ (full line).}
\label{fig_vis_noneq_angles}
\end{figure}%

We now turn to behaviour of the interferometer. We
consider first the case $t_a^2=1/2$
and $d_1=d_2$. Results for the visibility of AB fringes as a function of
bias voltage and interaction strength are presented in Fig. \ref%
{fig_visibility_range_ind}. Without interactions, visibility is
independent of bias, having a value fixed by the tunneling probabilities in the QPCs.
Interactions generate
a dependence of visibility on bias. Visibility at small bias is unaffected by interactions,
because in this regime there is only one extra
electron inside the interferometer at a time, but with increasing voltage the
visibility follows a sequence of lobes separated by zeros. The width in bias voltage 
of lobes is inversely
proportional to $\gamma$ for small $\gamma$ and saturates at a
value close to $\pi\hbar v_F/ed$ for $\gamma\gg1$. The phase of 
AB-fringes is independent of bias inside the lobes and jumps by $\pi$ at the
zeros of visibility. Both features, the lobes in the visibility and the phase
slips, match those of experiment by Neder \textit{et. al.} (see Figs.~2 and 3
of Ref.~\onlinecite{heiblum2}).

It is interesting to note that the visibility can become larger than one.
This signals a negative value of the differential conductance. 
Similar behavior has been
reported previously.\cite{chalker07}

Results for an interferometer with
unequal arm lengths, $d_2/d_1=1.2$, are presented in
Fig. \ref{fig_vis_nonequal_arms}. The gross behavior at intermediate and large interaction strengths is similar to that for an interferometer with equal length arms. The visibility minima for $d_1\not= d_2$, however, are not exact zeros: they approach zero at large interaction strength, but disappear altogether in the opposite limit of weak interactions. There is a corresponding evolution with interaction strength in the dependence of the phase of
AB fringes on bias voltage. In the absence of interactions
this phase varies linearly with bias for
an MZI with different length arms because the Fermi wavevector $k_F$ is linear in bias and the phase
difference between particles traversing the two arms is $k_F(d_2 - d_1)$.
With increasing interaction strength the phase dependence on bias develops
into a series of smooth steps, each of height $\pi$.
The risers of these steps coincide with minima of the visibility.
Strikingly, for the interacting system phase steps at minima of the visibility
persist for $d_1=d_2$, even though in this case phase would be independent
of bias without interactions.  Between risers the AB phase is almost independent of the ratio $d_2/d_1$ when interactions are strong, because of the cancellation between kinetic and interaction contributions to plasmon phase, as discussed above.
The stepwise phase variation we find at large
interaction strength also matches experimental observations (see Fig.~2 of Ref.~\onlinecite{heiblum2}).

Behavior of the visibility is insensitive to the transmission probability $%
t_b^2$ at QPC $b$, apart from the overall scale. Departures from $t_a^2=1/2$,
however, like unequal arm lengths, eliminate the exact zeros in visibility, leaving only sharp
minima. The dependence of visibility on voltage for
$t_a^2=0.75$ is shown in Fig. \ref{fig_vis_noneq_angles}.

\subsection{Exponential interaction}\label{sub:finite-range}

We next consider the interaction potential of Eq.~(\ref{exp_inter}), 
which decays exponentially with separation.
It is characterised by $g$, the interaction strength, and $1/\alpha$, the range.
In this case the
Schr\"odinger equation, Eq.~(\ref{eom-f}), for $0\leq x \leq d$ has the form 
\begin{equation}  \label{exp_int_eq}
-i\omega_p f_p(x)+v_F\partial_x f_p(x)=-\frac{g}{2\pi\hbar}%
\int_{0}^{d}e^{-\alpha|x-y|}\partial_{y}f_p(y)dy\,.
\end{equation}
Differentiating Eq. (\ref{exp_int_eq}) twice with respect to $x$
we obtain
\begin{multline}  \label{third_order_diff}
\partial_{xxx}f_p(x)-i(\omega_p/v_F)(\partial_{xx}f_p(x)-\alpha^2f_p(x)) \\
-(\alpha/d)(2\gamma+\alpha d)\partial_x f_p(x)=0
\end{multline}
with $\gamma=gd/(2\pi\hbar v_F)$ as before. This equation has a solution
of the form
\begin{equation}  \label{exp_sol}
f_p(x)=\sum_{n=1}^{3}A_n e^{i k_n x},
\end{equation}
where the $A_n$ are in general complex coefficients and the wavevectors $k_n$ are
obtained by solving the cubic equation 
\begin{equation}
\omega_p=k v_F[1+\tilde{U}(k)/2\pi\hbar v_F]\,.
\end{equation}
Here $\tilde{U}(k)=2\alpha g/(k^2+\alpha^2)$ is the Fourier transform of 
the potential in Eq.~(\ref{exp_inter}). 
Substitution of Eq.~(\ref{exp_sol}) into Eq.~(\ref{exp_int_eq}) 
gives two linear equations on $A_n$ and the
boundary condition $f_p(0)=1$ yields a third equation. These
determine the coefficients $A_n$ and read 
\begin{eqnarray}  \label{lin_sys_exp}
\sum_{n=1}^{3}A_n=1\,, \quad 
\sum_{n=1}^{3}A_n \frac{k_n e^{i k_n d}}{k_n+i\alpha}=0\,, \notag \\
\quad {\rm and} \qquad \sum_{n=1}^{3}A_n \frac{k_n}{k_n-i\alpha}=0 \,. 
\end{eqnarray}
Solving Eqns. (\ref{exp_sol}) and (\ref{lin_sys_exp}) numerically we
obtain the phase shifts from
\begin{equation}\label{ph_sh_exp}
\delta_p=-\mathrm{Arg}[f_p(d)e^{-i\omega d/v_F}].
\end{equation}
\begin{figure}[hbp]
\epsfig{file=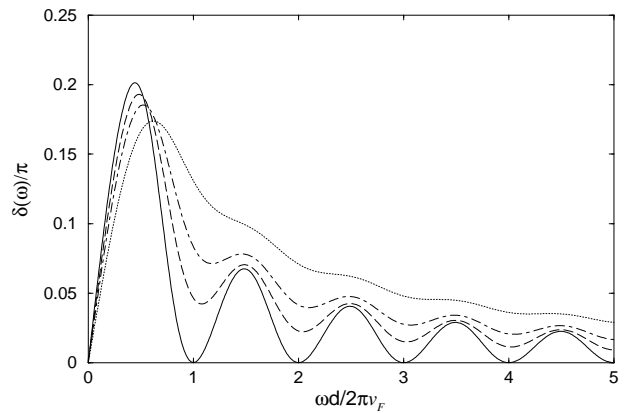,width=5.5cm,angle=270}
\caption{Frequency dependence of the plasmon phase shift for exponential
interactions of strength $\protect\gamma=0.5$ and
range: $\protect\alpha d=2$ (dotted line), $\protect\alpha d=1$ (dot-dashed
line), $\protect\alpha d=0.5$ (dashed line), and for a charging interaction, 
$\protect\alpha d=0$ (solid line).}
\label{fig_phase_shifts}
\end{figure}

The frequency dependence of this phase shift is shown in Fig. \ref{fig_phase_shifts} for different values of the interaction range $1/\alpha$ and the interaction strength $\gamma=1$. 
The main features, of a linear  variation at low frequency and a phase shift approaching zero at high frequency, are independent of range.

\begin{figure}[tbp]
\epsfig{file=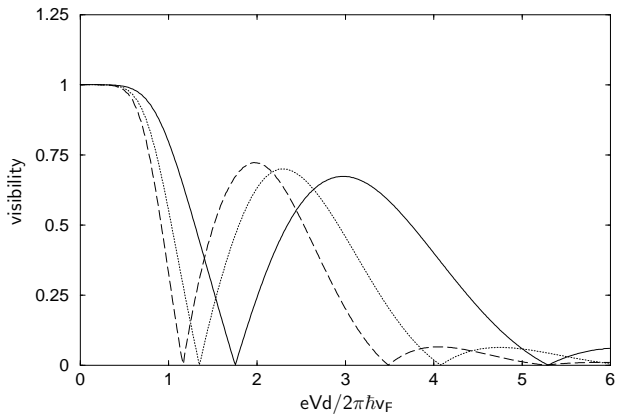,width=8cm,angle=0}
\caption{Visibility as a function of bias voltage for an MZI with
exponential interactions [Eq. (\protect\ref{exp_inter})],
with $d_{2}=d_{1}$, $t_{a}^{2}=t_{b}^{2}=1/2$ and
$\protect\gamma=0.5$, for interaction ranges: $\protect\alpha d=0.5$ (dashed line), $\protect\alpha d=1$
(dotted line), $\protect\alpha d=2$ (solid line).}
\label{fig_vis_finite_range}
\end{figure}

The resulting fringe visibility in an MZI with these interactions is illustrated in Fig. \ref{fig_vis_finite_range}.
At fixed interaction strength the visibility has zeros at values of the bias voltage
which are
set by the energy scale $\hbar v_F \alpha$
for interactions with range much shorter than the arm length 
and by   $2\pi\hbar v_F/d$ for interactions with range 
in the opposite limit.

\subsection{Coulomb interaction}\label{sub:coulomb-interactions}

\begin{figure}[tbp]
\epsfig{file=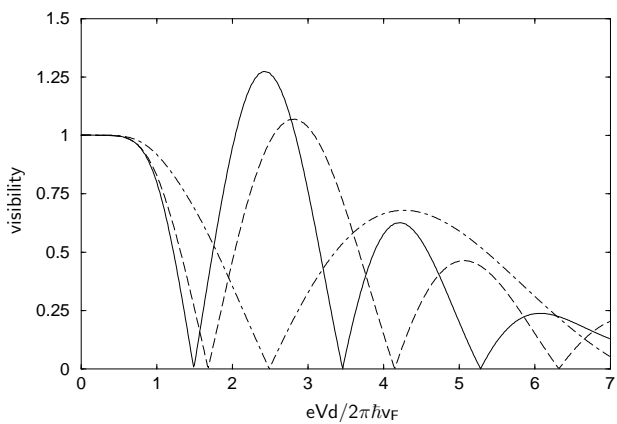,width=8cm,angle=0}
\caption{Visibility as a function of bias voltage for an MZI with 
unscreened Coulomb interactions [Eq. (\protect\ref{coulomb_int})],  
taking a short-distance cutoff $a_c=0.1 d_1$, 
$d_{2}=d_{1}$, $t_{a}^{2}=t_{b}^{2}=1/2$, and interaction strengths: $\protect%
\gamma_c=0.05$ (dot-dashed line), $\protect\gamma_c=0.1$ (dashed line) and $%
\protect\gamma_c=0.15$ (solid line).}
\label{fig_coulomb}
\end{figure}

Finally we consider the unscreened Coulomb interaction, Eq.~(\ref{coulomb_int}),
which is characterized by its strength $%
\gamma_c=g_c/2 \pi\hbar v_F$ and the short-distance cutoff $a_c$. We
treat the regime $a_c \ll d$; in the opposite limit it is similar to the charging
interaction discussed in Section \ref{sub:exact_gen_const_int}.
(Coulomb interactions have been studied previously,\cite{chalker07} without our restriction that they act just within the MZI, but only perturbatively in tunneling at QPCs.)
To calculate the plasmon phase shifts in this case we solve the
Bogoliubov equations [Eq.~(\ref{eq1AB})] numerically.
Results for the bias dependence of visibility 
are presented in Fig.~\ref{fig_coulomb}.
The visibility again shows lobes. Their width in bias voltage is inversely proportional to the interaction strength $\gamma_c$ for weak interactions and is set by the interferometer energy scale $2\pi\hbar v_F/d$ for strong
interactions. 

In summary, while the detailed shape of oscillations in visibility with bias voltage depend on the model used for interactions, the main features are independent of this choice.

\section{Discussion of the approach of Neder and Ginossar}\label{sec:discussion_neder}

In a recent paper\cite{neder08} Neder and Ginossar have presented an approximate treatment of interaction effects in an MZI, using the charging interaction of Eq.~\ref{pot_range_ind}.
They arrive at a result similar in form to the exact one given above in Eqns. (\ref{psi_dt}) 
and (\ref{phase-operator}), but with a different kernel $Q(x)$:
see the comparison in Fig. \ref{fig_kernel}. Making further simplifications in the course of a numerical evaluation, they obtain nodes in the dependence of visibility on bias voltage.
We discuss their approach in this section (we note that various choices of kernel and of approximations schemes in this type of numerical evaluation have also been discussed in Ref.~\onlinecite{neder+marquardt}). We show that their central approximation is equivalent to neglect of the chiral anomaly in the Tomonaga-Luttinger model, and explain physically what this entails. We also examine how far the differences between results for visibility in Ref.~\onlinecite{neder08}  and those in the present paper are due to this central approximation, and how far they stem from simplifications of the numerical evaluation made in Ref.~\onlinecite{neder08}. To do this we make a numerically exact calculation of the fringe visibility using the kernel of  Ref.~\onlinecite{neder08} and compare the outcome with our own results.

\subsection{Single edge with translationally invariant interactions}

The approach under discussion starts from the Heisenberg equation of motion for an operator $\hat{A}_H(t)$. Consider first a single edge channel without QPCs. 
In this case the equation of motion is
\begin{equation}
i\hbar\partial _{t}\hat {A}_H(t)=[\hat{A}_H(t),\hat{H_0}]
\label{eq_mot}
\end{equation}%
with initial condition $\hat{A}_H(0)=\hat A$, where $\hat A$
is the operator in the Schr\"odinger representation. 
In the absence of interactions Eq.~(\ref{eq_mot}) for the field operator has the solution
\begin{equation}
\hat{\psi}_{H}^{+}(d_{},t)=\hat{\psi}_{H}^{+}(\varepsilon
,t-d_{\eta}/v_{F})\,.
\end{equation}
Next, include translationally invariant
interactions $U(x,x^{\prime})=U(x-x^{\prime})$, with $U(0)=0$ to avoid
self-interactions. The equation of motion is
\begin{multline}  \label{eq_mot_int}
i\hbar (\partial _{t}+v_{F}\partial _{x})\hat{\psi}^{+}_{H}(x,t)=\\-
\hat{\psi}^{+}_{H}(x,t)\int_{-\infty }^{\infty }U(x-x^{\prime})\hat\rho_{H}
(x^{\prime},t)dx^{\prime}.
\end{multline}%
It  apparently has the solution 
\begin{equation}
\hat{\psi}^{+}_{H}(x,t)=e^{i\hat{\theta}_{H}(x,t)}\hat{\psi}%
^{+}_{H}(x-v_{F}t,0)  \label{lieb_trans}
\end{equation}%
where the phase operator $\hat\theta_{H} (x,t)$ is given by
\begin{equation}  \label{gauge_trans}
\hat{\theta}_{H}(x,t)=\frac{t}{\hbar}\int_{-\infty }^{\infty }U(x^{\prime})%
\hat{\rho}(x-v_Ft+x^{\prime},0)dx^{\prime}.
\end{equation}

This calculation is essentially a version in the Heisenberg picture of the solution
of the Tomonaga-Luttinger model by Luttinger,\cite{luttinger} who used a canonical transformation to diagonalise the Hamiltonian. It is exact provided the number of particles in the system is finite. The difficulty, of course, is that for finite particle number a model with linear dispersion has no ground state, and so one wants to introduce a filled Fermi sea. Unfortunately, as shown by Leib and Mattis,\cite{mattis}  the calculation is then no longer exact, because density operators which commute when particle number is finite no longer do so in the presence of the Fermi sea.

To illustrate the difference in physical behaviour between a system with a finite number of particles and one with a Fermi sea, it is useful to consider the time evolution of an initial state in which single particle orbitals  are all occupied for wavevectors $k$ in the range $0\leq k\leq k_{\mathrm{F}}$, and others are all empty.  We denote this state by $|k_F\rangle$ and compare its evolution with that of the state with a Fermi sea, in which {\em all} orbitals with $k\leq k_F$ are occupied. The latter is an exact eigenstate of ${\hat H}_0$ for arbitrary choice of interaction, and so has trivial time evolution. By contrast, $|k_F\rangle$ is not an eigenstate in the presence of interactions. To be specific, we calculate the time dependence of the particle number $\hat{n}(Q)=\hat{c}_{Q}^{+}\hat{c}_{Q}$ in an orbital with wavevector $Q$. Within a short time expansion we have
\begin{equation}
\langle \hat{n}(Q,t)\rangle =\langle \hat{n}(Q)\rangle +\mathrm{i}t\langle \lbrack 
\hat{H}_0,\hat{n}_{Q}]\rangle -\frac{t^{2}}{2}\langle \lbrack \hat{H}_0,[%
\hat{H}_0,\hat{n}_{Q}]]\rangle \ldots
\end{equation}%
For the state $|k_{\mathrm{F}}\rangle $ the average $\langle \lbrack 
\hat{H}_0,\hat{n}(Q)]\rangle =0$. The second order term is conveniently expressed in terms of
\begin{equation}
\tilde{U}(q)=\int_{-\infty }^{\infty }dx\,U(x)e^{iqx}\,,
\end{equation}%
the Fourier transform of the interaction potential.
Taking $k^{-1}_{\mathrm{F}}$ to be much smaller than the interaction range, and
writing $Q=k_{\mathrm{F}}+P$ with $P>0$, we find
\begin{eqnarray}
\langle \hat{n}(Q,t)\rangle &=&t^{2}\langle \hat{H}_0 \hat{n}(Q)\hat{H}_0\rangle
+\ldots  \notag \\
&=&\frac{t^{2}}{8\pi ^{2}}\int_{P}^{\infty }\mathrm{d}q\,q|\tilde{U}(q)|^{2}+%
\mathcal{O}(t^{3})\;.  \notag
\end{eqnarray}%
Similarly, for $-k_F \ll P<0$ we obtain
\begin{equation}\label{decay}
\langle \hat{n}(Q,t)\rangle =1-\frac{t^{2}}{8\pi ^{2}}\int_{-P}^{\infty }\mathrm{d}%
q\,q|\tilde{U}(q)|^{2}+\mathcal{O}(t^{3})\;.
\end{equation}
This shows that interactions scatter particles from the orbitals they occupy initially in $|k_F\rangle$ to others of larger and smaller energy. The pair of particles involved in a typical scattering event has one initial wavevector just less than $k_F$ and the other just larger than zero. One particle is scattered to a state with wavevector $k > k_F$  and the other to a state with negative wavevector. The rate for this process remains finite even when $k_F$ is large, but in the presence of a Fermi sea these scattering processes are blocked by Pauli exclusion. 

\subsection{Interactions confined to a finite region}

To demonstrate that no other approximations are involved in the derivation of the kernel of Ref.~\onlinecite{neder08}, we next set out a calculation equivalent to the one leading to Eq.~(\ref{lieb_trans}), but for 
interactions that operate only in the finite
region $0<x<d $ and are translationally invariant within this
region, so that 
\begin{equation}
\mathcal{\hat{H}}_{int}=\frac{1}{2}\int_{0}^{d}\int_{0}^{d}U(x-x^{\prime})%
\hat{\rho}(x)\hat{\rho}(x^{\prime})dxdx^{\prime}\,.
\end{equation}%
Then the equation of motion for $\hat{\psi}^{+}(x,t)$ 
can be solved by integrating forwards in time separately in each of three regions 
and matching at boundaries.
For $x<0$ we have
\begin{equation}
\hat{\psi}^{+}(x,t)=\hat{\psi}^{+}(x-v_F t,0).
\end{equation}
Within the approximation under discussion, for $0<x<d$
\begin{equation}
\hat{\psi}^{+}_{H}(x,t)=e^{i\hat{\theta}_{H}(x,t)}\hat{\psi}%
^{+}_{H}(x-v_{F}t,0),
\end{equation}%
where $\hat\theta_{H}(x,t)$ satisfies the equation 
\begin{equation}
(\partial _{t}+v_F\partial _{x})\hat\theta_H(x,t)=\frac{1}{\hbar}%
\int_{0}^{d}U(x-x^{\prime})\hat{\rho}_{H}(x^{\prime},t)dx^{\prime}
\end{equation}%
with the boundary condition $\hat{\theta}(0,t)=0$.
This has the solution 
\begin{multline}
\hat\theta_{H}(x,t)=\frac{1}{\hbar v_F}\left\{
\int_{-v_{F}t}^{x-v_{F}t}(x'+v_{F}t)+ \right. \\
\left. x\int_{x-v_{F}t}^{d-v_{F}t}
+\int_{d-v_{F}t}^{d+x-v_F t} (d+\tilde x)\right\}\times\\
U(\tilde x)\hat{\rho}_{H}(x',0)dx',
\end{multline}%
in which $\tilde x=x-v_{F}t-x'$.

In particlular, for $x=d$ we have 
\begin{multline}\label{theta}
\hat{\theta}_{H}(d,t)=\frac{1}{\hbar v_{F}}\int_{0}^{d}x^{\prime}U(d-x^{%
\prime})\hat{\rho}_{H}(x^{\prime},t)dy \\
+\frac{1}{\hbar v_{F}}\int_{d}^{2d}(2d-x^{\prime})U(d-x^{\prime})\hat{\rho}%
_{H}(x^{\prime},t)dx^{\prime}\,.
\end{multline}%
In the region $x>d$ we can use results from $0<x<d$ as a
boundary condition to obtain 
\begin{equation}\label{lutt}
\hat{\psi}^{+}_{H}(x,t)=e^{i\hat{\theta}_{H}(d,t)}\hat{\psi}%
^{+}_{H}(x-v_{F}t,0)\,.
\end{equation}
For the charging interaction, Eq.~(\ref{pot_range_ind}), this gives an expression
for the kernel $Q(x)$ appearing in Eq.~(\ref{phase-operator}) which is equivalent to Eq.~(11) of Ref.~\onlinecite{neder08}.

\subsection{Interferometer with internal interactions}

\begin{figure}[tbp]
%[htb]
\epsfig{file=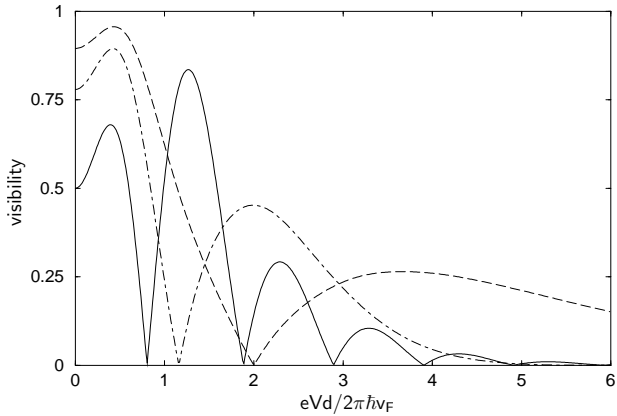,width=8cm,angle=0}
\caption{Visibility as a function of bias voltage for an MZI with charging interactions, calculated using the kernel of Eq. (11)
of Ref. \protect\onlinecite{neder08}, with
$d_1=d_2=1$ and $t_{a}^{2}=t_{b}^{2}=1/2$, for interaction
strengths: $\protect\gamma=0.1$ (dashed line), $\protect\gamma=0.2$
(dot-dashed line) and $\protect\gamma=0.5$ (solid line).}
\label{fig_neder}
\end{figure}

For an interferometer with only internal interactions, the foregoing discussion 
can be combined with the approach described in Sections \ref{bosonization}, \ref{scattering} and \ref{sub:exact_diag_ham}. In particular, making use of Eq.~(\ref{lutt}) in place of Eq.~(\ref{kernel_mom}), one arrives at alternative results for visibility.
We have evaluated these without further approximation for charging interactions.
They are displayed in Fig.~\ref{fig_neder}. In this approximation the visibility as a function of bias shows a lobe pattern similar to that given by our exact solution (Fig.~\ref{fig_visibility_range_ind}) but some significant differences are apparent. In particular, within the approximate treatment visibility at small bias is reduced from its value in the non-interacting system, and more so as interaction strength is increased. This loss of coherence mirrors the scattering described by Eq.~(\ref{decay}). It is not a feature of the exact treatment for a system with a Fermi sea: in that case at small bias extra electrons are dilute and pass independently through the interferometer. 

\section{Perturbation theory in interactions}\label{sec:pert-theory}

The approach we have used to obtain exact results can be applied only to models
in which interactions act solely within the MZI. In order to gain some
understanding of the consequences of more general interactions, it is useful to
formulate perturbation theory in powers of interaction strength. We outline such a calculation in this section. Since details are quite messy (around forty separate terms appear at first order in interaction strength), we limit ourselves to sketching how a new physical effect -- a second harmonic in the variation of differential conductance with AB flux -- appears when there are interactions between the interior and exterior of the MZI.  This expansion in powers of interaction strength, which can be applied for arbitrary values of the tunneling amplitude at QPCs, is complementary to the perturbation theory in tunneling strength, developed for arbitrary interaction strength in Ref.~\onlinecite{chalker07}.

We separate the Hamiltonian, Eq.~(\ref{ham}), into the single-particle contribution 
$\hat{H}_{1}=\hat{\mathcal{H}}_{kin}+\hat{\mathcal{H}}_{tun}$ and the interaction term
$\hat{\mathcal{H}}_{int}$, and start from a single-particle basis of scattering states of $\hat{H}_{1}$. These states are labelled by energy $\hbar v_F q$ and by the channel $\eta$ (denoted 1 and 2 in Fig.~\ref{fig2}) from which particles are incident on the MZI. They have amplitude $e^{iqx}\varphi_{q\eta}(x,\kappa)$ at point $x$ in channel $\kappa$. Thus, for example, 
$\varphi_{q1}(x,\kappa)$ takes the form given in Table 1.
\begin{table}[h]
\begin{tabular}{|c||c|c|c|}\hline
$\kappa$ & $x<0$ & $0 < x < d_\kappa$ & $d_\kappa < x$ \\ \hline \hline
1 &  1 & $r_a$ & $r_a r_b - t_a t_b e^{i(q[d_2 - d_1]+\beta - \alpha)}$\\ \hline
2 &   0 & $-it_a e^{-i\alpha} $ & $- i t_a r_b e^{-i\alpha} - i r_a t_b e^{i(q[d_1 - d_2]-\beta)}$\\ \hline
\end{tabular}
\caption{Amplitudes of the scattering state $\varphi_{q1}(x,\kappa)$.}
\end{table}

We consider an initial many-particle wavefunction $|\Psi_0\rangle$ in which scattering states are  occupied up to Fermi wavevectors $p_1$ and $p_2$ respectively in the two incident channels, representing a bias voltage $V = \hbar v_{F}(p_1-p_2)/e$. Using the interaction representation 
$\hat{A}\left( t\right) =e^{i\hat{H}_{1}t/\hbar }\hat{A}e^{-i\hat{H}_{1}t/\hbar } $
based on $\hat{H}_{1}$ (different, of course, from that employed in Section \ref{sec:exact-solution}), we evolve the initial state forward in time from $t=-\infty$ to $t=0$. The current at zeroth and first order in $\hat{\mathcal{H}}_{int}$ is 

\begin{multline}
\langle \hat{I}\left( t\right) \rangle _{t=0}
=\left\langle \Psi _{0}\right\vert \hat{I}\left(
0\right) \left\vert \Psi _{0}\right\rangle  \label{ref_cur1} \\
+\frac{i}{\hbar }\int_{-\infty }^{0}\left\langle \Psi _{0}\right\vert \left[ 
\hat{\mathcal{H}}_{int}\left( \tau \right) ,\hat{I}\left( 0\right) \right] \left\vert
\Psi _{0}\right\rangle d\tau.
\end{multline}

To evaluate the interaction term in this expression we take matrix elements
in the basis of single-particle scattering states, denoting 
for brevity the pair of labels $q_a,\kappa_a$ by $a$. For an interaction $U(x,x^\prime)$ that does not couple channels we write
\begin{multline}
 \bar U_{abcd}=\frac{1}{2}\int\int {\rm d}x {\rm d}x'\,U(x,x')\,e^{i[(q_d-q_a)x+(q_c-q_b)x^\prime]}\\
\times \sum_\kappa \varphi_{q_a \kappa}^{*}(x)\varphi_{q_b\kappa}^{*}(x')\varphi_{q_c\kappa}(x')\varphi_{q_d\kappa}(x) \label{U_{abcd}}
\end{multline}
and define the antisymmetrised combination $U_{abbd}=\frac{1}{2}(\bar U_{abbd}-\bar U_{abdb})$. We evaluate matrix elements $I_{ab}$ of the current operator by considering in the first instance a tunneling contact of finite width $w$, then taking the limit $w\to 0$, as in Section \ref{sub:exact_cur_der}. For example,
\begin{multline}
I_{k1\,q1} = ev_F\left(t_a^2 + r_a^2t_b^2 e^{id_1(q-k)} - t_a^2 t_b^2 e^{id_2(q-k)}\right. \\
\left.+t_ar_at_br_b\left[e^{i(\alpha-\beta+qd_1-kd_2)} + e^{i(\beta-\alpha+qd_2-kd_1)}\right]\right)\,.\nonumber
\end{multline}
 
With this notation, and denoting by $n_a$ the average occupation of the state $a$, 
the first order term from Eq.~(\ref{ref_cur1})  has the form
\begin{multline}
\frac{i}{2 \pi^3 \hbar }\int_{-\infty
}^{0}\!\!\!{\rm d}\tau \int_{-\infty}^{\infty}\!\!\! {\rm d}k_a \int_{-\infty}^{\infty}\!\!\!{\rm d}k_b 
\int_{-\infty}^{\infty}\!\!\!{\rm d}k_c\\
\times U_{abbc}I_{ca}n_{b}\left( n_{a}-n_{c}\right) e^{iv_F\left(
k_{a}-k_{c}\right) \tau }\,.  \nonumber \label{current}
\end{multline}
Evaluating the integrals on $\tau$, $k_a$, $k_b$ and $k_c$ we obtain an expression in which only integration on $x$ and $x^\prime$ from Eq.~(\ref{U_{abcd}}) remains. Because the factors $\varphi_{q\eta}(x,\kappa)$ take different forms in each of the regions I, II and III, defined in Fig.~\ref{fig2}, these final integrals naturally separate into distinct contributions according to the possible locations of each of two interacting particles. 

As an illustration of this general approach we consider the contribution to current arising at first order from the interaction between a particle in region II and one in region III. This is made up of an exchange term $\delta I_F^{23}$ and a Hartree term   $\delta I_H^{23}$. To write expressions for these in a concise form we limit ourselves to the case $d_1=d_2\equiv d$ and
define the constant $\mathcal{T}_{ab}=t_a t_b r_a r_b$. Then
\begin{multline}
\delta I_{F}^{23}=-\frac{8e\mathcal{T}_{ab}^2}{\pi \hbar} \sin 2\Phi \\ \times
\int_{0}^{d} {\rm d}x\,\int_{d}^{\infty}
{\rm d}x^{\prime}\,
U(x,x^{\prime})
\frac{\sin^2[(p_1-p_2)(x-x^\prime)/2]}{(x-x^\prime)^2}\nonumber
\end{multline}
and 
\begin{multline}
\delta I_{H}^{23}=\frac{2e\mathcal{T}_{ab}^2}{\pi \hbar} (p_1 - p_2)^2\sin 2\Phi  \\
\times \int_{0}^{d}{\rm d}x\int_{d}^{\infty}{\rm d} x^\prime \, U(x,x^{\prime})
\,.\nonumber
\end{multline}
As advertised, both terms involve the second harmonic $\sin 2 \Phi$ of the AB phase $\Phi$.
They  display the antisymmetry expected for the current under reversal of both $V$ and $\Phi$. 
A second harmonic is also produced by the interactions between the electrons in the
regions I and III which we do not present here. By contrast, a similar perturbative treatment of interactions between electrons in region II generates only zeroth and first  harmonics of $\Phi$, as expected from the results presented in Section \ref{sec:exact-solution}.

\section{Discussion}\label{sec:conclusion}

In summary, we have argued that electron interactions are the 
origin of the experimentally observed dependence of visibility of 
AB oscillations on bias voltage in electronic MZIs. 
This is illustrated most simply by calculations of two-particle interference 
effects, described in Section \ref{sec:toy-model}, and demonstrated in 
detail by exact results for the many body system, presented in Section \ref{sec:application_potentials}. 
Our calculations rely on a simplified form for interactions, but we believe
our choice is quite reasonable. Our central approximation is to neglect
interactions between an electron inside the MZI and one outside. In
practice, such interactions will anyway be screened by the metal gates that
define the QPCs. We also neglect interactions between a pair of electrons
that are both outside the MZI. This is unimportant: before electrons reach
the MZI, such interactions do not cause scattering because of Fermi
statistics, while after electrons pass through the MZI, these interactions
cannot affect the current. We have given results for three forms of interaction within
the MZI, finding their main features to be independent of details of the model.
These features include a series of lobes of decreasing amplitude 
in AB fringe visibility as a function of bias, with jumps by $\pi$ in the phase of 
AB oscillations near minima in visibility. 

The width in bias voltage of the central visibility lobe defines an energy
scale. For our model of charging interactions this scale is of order $g$ at large $\gamma$. Taking $v_F = 2.5\times 10^{4} \mathrm{ms^{-1}}$, $d=10\mu\mathrm{m}$ and the
permittivity $\epsilon =12.5$ of GaAs, we estimate from the capacitance of
an edge channel $g\sim 10\mathrm{\mu eV}$. This is similar to the
experimentally observed value of about $14\ \mathrm{\mu eV}$, given in 
Ref.~\onlinecite{heiblum2}.

We close by comparing the results we have presented with those from other approaches.
As a first step, we note that, while the existence of multiple side lobes in the visibility of AB oscillations as a function of bias voltage cannot be accounted for by a simple treatment of dephasing, a single side lobe can emerge from a simple, phenomenological treatment. To see this, consider the dependence of current $I(V,\Phi)$ on bias $V$ and AB flux $\Phi$. Assuming that there is a just one harmonic in AB oscillations, we have
\begin{equation}
I(V,\Phi) = I_0(V) + I_{1}(V) \cos(\Phi + \theta(V))\,.
\end{equation}
With this notation, the visibility is
\begin{equation}\label{dephasing}
{\cal V} = \frac{[(\partial_V I_{1}(V))^2 + (I_1(V) \partial_V \theta(V))^2]^{1/2}}{|\partial_V I_0(V)|}\,.
\end{equation}
It is reasonable to expect quite generally that $I_1(V)$ increases with $V$ for small $V$, has a single maximum, and decreases towards zero at large $V$, so that $\partial_V I_1(V)$ has a single zero for $0\leq V<\infty$. If in addition the phase $\theta(V)$ of AB oscillations is independent of $V$ (and only in this case), the existence of a single side lobe in visibility follows. 

The theoretical difficulty, then, is to understand the observation of multiple side lobes. Calculations to date that generate such behaviour can be divided into three categories. One of these\cite{sukhorukov07} involves a plasmon resonance between one arm of the MZI and a counter-propagating edge state at the boundary of the Hall bar. Such a coupling of the MZI to another edge state is not an essential part of the interferometer design, and in this sense the mechanism is not an intrinsic one. For that reason, it seems unlikely to provide the explanation for observations in many different samples of varying designs. The most important comparison is therefore between a second category of explanation,\cite{sukhorukov08} which is based on coupling between the two channels existing at each edge for filling factor $\nu=2$, and the third category, which is formed by the calculations that we have presented, together with earlier approximate discussions \cite{neder08,sim08} of similar physics, and has been worked out for a system at $\nu=1$. 

According to the approach of Ref.~\onlinecite{sukhorukov08}, multiple side lobes should be found only at $\nu=2$, which seems indeed to be the case experimentally.
Some important discrepancies between this theory and experiment remain, however. One is that, within the approach of Ref.~\onlinecite{sukhorukov08}, the shape of the envelope of the lobe pattern is controlled by the difference in interferometer arm lengths: in particular, for an interferometer with arms of equal length, visibility does not fall to zero at large bias. This is in conflict with observations. It requires one to assume\cite{sukhorukov08} that two separate physical processes are involved: the process included in the theory, which leads to multiple zeros in visibility at $\nu=2$, and another one, omitted from the theory (such as dispersion of the edge modes), which controls decay of the envelope. Moreover, if this theory is adopted at $\nu=2$, the existence of a single side lobe at $\nu=1$ must be attributed to a separate dephasing mechanism, following Eq.~(\ref{dephasing}). 
In experiment, there appears to be a common voltage scale determining  all aspects: the zeros in visibility at $\nu=2$, the envelope of the lobe pattern at either filling factor, and the position of the visibility zero at $\nu=1$. It would be a surprising coincidence if two separate mechanisms were both to involve the same scale. An explanation of multiple side lobes in visibility based exclusively on dispersionless slow and fast edge modes at $\nu=2$ therefore seems problematic.
By contrast, the calculations we have presented generate zeros of visibility and an overall decaying envelope for visibility from a single mechanism. Depending on interaction strength we find either multiple side lobes or only a single prominent side lobe. The mechanism is essentially plasmon dispersion. It is likely that a full understanding of experiments will require a combination of both aspects -- dispersion and the existence of two modes at $\nu=2$. It is however, an important point of principle, demonstrated by the calculations we have described, that multiple side lobes are not an exclusive consequence of coupling between a slow and a fast mode at $\nu=2$.

A definite qualitative feature of our results is that exact zeros of fringe visibility
at certain values of bias voltage appear only for an interferometer with equal length arms, 
in which the transmission probability at the first QPC is tuned precisely to the value one half. 
Changes in transmission probability and (though to a much lesser extent, if interactions are strong) arm length from these values convert exact zeros to finite minima in visibility. Such sensitivity to transmission probability has not been reported experimentally. The reason it appears in our theory can be understood starting from Eq.~(\ref{dephasing}): exact zeros of visibility can occur only if the phase $\theta(V)$ of AB oscillations is independent of bias $V$.
A phase independent of bias is ensured by symmetry for an interferometer with $L_1=L_2$ and $t_a^2=1/2$. For other parameter choices, our model, in which there are no interactions between electrons on opposite interferometer arms, yields a bias-dependent $\theta(V)$, essentially because that Hartree potentials for electrons on each arm vary differently with $V$. For this reason, it would be interesting to generalise our treatment to systems with interactions between edges as well as within each edge. We expect that such interactions would reduce or eliminate the dependence of $\theta(V)$ on $V$, giving near or exact zeros of visibility even for $t_a^2\not= 1/2$ or $L_1\not= L_2$, in accord with experiment. Other possible generalisations include calculation of finite temperature effects and of noise power.

\section{Acknowledgements}

We thank V. V. Cheianov, F. H. L. Essler, Y. Gefen, D. Mailly, F. Pierre and P. Roche for fruitful discussions, and acknowledge support
from EPSRC grants EP/D066379/1 and EP/D050952/1.

\appendix
\section{Calculation of the commutation relations}\label{app:comm_rel}

We wish to show that the commutators $[\hat{\mathcal{H}}_{tun}^{a}(t_1),\hat{\mathcal{H}}
_{tun}^{b}(t_2)]$ and $[\hat{G}_{12}(t_1),\hat{\mathcal{H}}_{tun}^{b}(t_2)]$ appearing in Section \ref{sub:exact_smat_eval} are zero for $t_1\geq  t_2$. As a first step we
obtain commutation relations for the fermion fields
appearing in  Eq.~(\ref{psi_dt}) of section \ref{sub:exact_eval_corr}. To keep notation concise, we define
\begin{equation*}
\hat{\Psi}_{\eta }\left( z_\eta\right) \equiv \hat{\psi}_{\eta }(d_{\eta
},t)=e^{-i\int_{-\infty }^{\infty }Q_{\eta }\left(x-z_\eta\right) \hat{\rho}%
_{\eta }\left( x\right) dx}\hat{\psi}_{\eta }\left(z_\eta\right),
\end{equation*}%
where $z_\eta=d_\eta-v_F t$.
The kernel $Q_{\eta }\left( z\right) $ is nonzero only for $z>-d_{\eta }$. This follows from causality (with interactions occurring only inside the interferometer, the phase of an electron at QPC $b$ cannot be influenced by electrons that have yet to enter the MZI), and can be demonstrated explicitly for the case of charging interactions, using the analytic structure of the phase shifts, Eq.~(\ref{phase_shifts}). We have
\begin{equation}
\{\hat{\psi}_{\eta }\left( x\right) ,\hat{\Psi}_{\eta ^{\prime }}\left(
x^{\prime }\right) \}=0,\text{\ }\{\hat{\psi}_{\eta }\left( x\right) ,\hat{%
\Psi}_{\eta ^{\prime }}^{+}\left( x^{\prime }\right) \}=0
\label{phi_psi_comm}
\end{equation}%
for any $\eta ,\eta ^{\prime }$ and $x^{\prime }-x\geq d_{\eta }$.
From this we find
\begin{multline*}
\lbrack \hat{\mathcal{H}}_{tun}^{a}(t_1),\hat{H}_{tun}^{b}(t_2)]= \\
\lbrack v_{a}e^{i\alpha }\hat{\psi}_{1}^{+}(-v_{F}t_1)\hat{\psi}_{2}(-v_{F}t_1)+
\mathrm{h.c.,} \\
v_{b}e^{i\beta }\hat{\Psi}_{1}^{+}(d_{1}-v_{F}t_2)\hat{\Psi}%
_{2}(d_{2}-v_{F}t_2)+\mathrm{h.c.]}
=0
\end{multline*}%
for $t_1 \geq t_2$.

We also require the anticommutators of $\hat\Psi _{\eta }\left( x\right) $ 
with itself and with $\hat\Psi^+_{\eta }\left( x\right) $. These are straightforward to
obtain starting from the bosonized form for the fermion operators. However, it is instructive also to derive them directly from the representation in Eq.~(\ref{psi_dt}), as follows.
We have
\begin{multline*}
\hat\Psi _{\eta }\left( x\right) \hat\Psi_{\eta ^{\prime }}\left( x^{\prime }\right)
=-\hat\Psi _{\eta ^{\prime }}\left( x^{\prime }\right) \hat\Psi _{\eta }\left(
x\right) \times \\
e^{-[\hat{Q}_{\eta },\hat{Q}_{\eta ^{\prime }}]}e^{i[Q_{\eta }\left(
x^{\prime }-x\right) -Q_{\eta }\left( x-x^{\prime }\right) ]\delta _{\eta
\eta ^{\prime }}},
\end{multline*}%
where%
\begin{multline*}
\lbrack \hat{Q}_{\eta },\hat{Q}_{\eta ^{\prime }}]\equiv \int_{-\infty}^{\infty} Q_{\eta
}\left( z-x\right) Q_{\eta ^{\prime }}\left( z'-x^{\prime }\right) \times \\
\lbrack \hat{\rho}_{\eta }\left( z\right) ,\hat{\rho}_{\eta ^{\prime
}}\left( z^{\prime }\right) ]dzdz^{\prime }.
\end{multline*}%
Because of the presence of the filled Fermi sea, the density operators
do not commute.\cite{vonDelft} Instead
\begin{equation*}
\lbrack \hat{\rho}_{\eta }\left( z\right) ,\hat{\rho}_{\eta ^{\prime
}}\left( y\right) ]=-\frac{i}{2\pi }\partial _{z}\delta \left( z-y\right)
\delta _{\eta \eta ^{\prime }}.
\end{equation*}%
From this
\begin{equation*}
\lbrack \hat{Q}_{\eta },\hat{Q}_{\eta ^{\prime }}]=\frac{i}{2\pi }%
\int_{-\infty }^{\infty } \!\!\!{\rm d}y\,Q\left( y-x^{\prime
}\right) \partial _{y}Q\left( y-x\right)\,.
\end{equation*}%
We now prove 
%for any $x,x^{\prime }$ 
that
\begin{multline*}
Q\left( x^{\prime }-x\right) -Q\left( x-x^{\prime }\right) = \\
\frac{1}{2\pi }\int_{-\infty }^{+\infty }\!\!\!{\rm d}y\,
Q\left( y-x^{\prime }\right) \partial _{y}Q\left( y-x\right).
\end{multline*}%
Introducing the Fourier transform of the kernel
we have%
\begin{multline*}
Q\left( x^{\prime }-x\right) -Q\left( x-x^{\prime }\right) = \\
\int {\rm d}q\, \tilde{Q}\left( q\right) [e^{iq\left(
x^{\prime }-x\right) }-e^{iq\left( x-x^{\prime }\right) }]
\end{multline*}
and
\begin{multline*}
\frac{1}{2\pi }\int_{-\infty }^{+\infty }\!\!\! {\rm d}y\,
Q\left( y-x^{\prime }\right) \partial _{y}Q\left( y-x\right)= \\
i \int {\rm d}q\,q\tilde{Q}\left( q\right) 
\tilde{Q}\left( -q\right) e^{iq\left( x^{\prime }-x\right) }.
\end{multline*}%
Thus we wish to show
\begin{equation*}
\tilde{Q}\left( q\right) -\tilde{Q}(-q)= i q\tilde{Q}\left(
q\right) \tilde{Q}\left( -q\right) .
\end{equation*}%
It is easy to check that this is the case, using the explicit form
\begin{equation*}
\tilde{Q}\left( q\right) =2\pi \gamma dj_{0}^{2}(qd/2)(1+\gamma
e^{-iqd/2}j_{0}(qd/2))^{-1}.
\end{equation*}
for the kernel in the case of a charging interaction, where $j_0(x) = (\sin x)/x$.
This gives 
\begin{eqnarray*}
\{\hat{\Psi}_{\eta }\left( x\right) ,\hat{\Psi}_{\eta ^{\prime }}\left(
x^{\prime }\right) \}&=&0, \nonumber \\ \{\hat{\Psi}_{\eta }\left( x\right),\hat{\Psi}
_{\eta ^{\prime }}^{+}\left( x^{\prime }\right) \}&=&\delta _{\eta \eta
^{\prime }}\delta \left( x-x^{\prime }\right). \nonumber  \label{phi_phi_comm}
\end{eqnarray*}
It follows directly that $[\hat{G}_{12}(t),\hat{\mathcal{H}}_{tun}^{b}(t^{\prime
})]=0$ for $t\geq t^{\prime }$.

\section{Calculation of the $S-$matrix}\label{app:s-matrix}

We require an explicit form for the unitary transformation of fermion
operators generated by
\begin{equation}
\hat{S}^{a}(t)=\mathrm{T}\exp \left\{ -\frac{i}{\hbar }\int_{0}^{t}\hat{\mathcal{H}}
_{tun}^{a}\left( \tau \right) d\tau \right\}\,.  \label{SA_I}
\end{equation}
The fermion operators appearing in $\hat{\mathcal{H}}
_{tun}^{a}$ commute with $\hat{\mathcal{H}}_{int}$ because their
position coordinates are before the interacting region. For this reason
the interaction representation of $\hat{\mathcal{H}}%
_{tun}^{a}(\tau )$ has the simple form
\begin{equation*}
\hat{\mathcal{H}}_{tun}^{a}\left( \tau \right) =v_{a}e^{i\alpha }\hat{\psi}%
_{1}^{+}\left( -v_{F}\tau \right) \psi _{2}\left( -v_{F}\tau \right) +%
\mathrm{h.c}.
\end{equation*}%
The Schr\"odinger operators $\hat{\psi}_{\eta }\left( -v_{F}\tau \right)$
anticommute at different values of their argument. As a consequence
\begin{equation*}
\lbrack \hat{\mathcal{H}}_{tun}^{a}\left( \tau \right) ,\hat{\mathcal{H}}_{tun}^{a}(\tau
^{\prime })]=0,
\end{equation*}%
for any $\tau ,\tau ^{\prime }$ and the time-ordering in (\ref{SA_I}) can be
omitted.

Defining
\begin{equation}
\tilde{\hat{\psi}}_{\eta }\left( x,t\right) =\hat{S}^{a+}(t)\hat{\psi}_{\eta
}\left( x\right) \hat{S}^{a}(t).
\end{equation}%
and using a Baker-Hausdorff formula%
\begin{multline*}
e^{-B}Ae^{B}=\sum_{n=0}^{\infty }\frac{1}{n!}[A,B]_{n} \\
=A+[A,B]+\frac{1}{2!}[[A,B],B]+...
\end{multline*}%
we obtain
\begin{equation}
\tilde{\hat{\psi}}_{\eta }\left( x,t\right) =\left\{ 
\begin{array}{c}
\sum_{\mu =1,2}\mathcal{S}_{\eta \mu }^{a}\hat{\psi}_{\mu }\left( x\right)
,\ \ -v_{F}t\leq x<+0 \\ 
\hat{\psi}_{\eta }\left( x\right) ,\ \ \mathrm{otherwise}%
\end{array}%
\right.  \label{psi_expl_A}
\end{equation}%
where
\begin{equation*}
\mathcal{S}^{a}=\left( 
\begin{array}{cc}
r_{a} & -it_{a}e^{i\alpha } \\ 
-it_{a}e^{-i\alpha } & r_{a}%
\end{array}%
\right) .
\end{equation*}%

\section{Matrix elements of the exponential operator}\label{app:mat_els}

Here we present a derivation of the equation which we use to
evaluate matrix elements of the form 
\begin{equation}
C_{kl}\equiv \langle \alpha |\hat{c}_{k}^{+}e^{i\sum_{ij}M_{ij}\hat{c}%
_{i}^{+}\hat{c}_{j}}\hat{c}_{l}|\beta \rangle,   \label{matel1}
\end{equation}%
where the fermionic operators $\hat{c}_{i}$ obey usual anticommutation
relations $\{\hat{c}_{i},\hat{c}_{j}^{+}\}=\delta _{ij}$ and $\{\hat{c}_{i},%
\hat{c}_{j}\}=0$. The matrix $M_{ij}$ is Hermitian and so the exponential 
\begin{equation}
\hat{U}=e^{i\sum_{ij}M_{ij}\hat{c}_{i}^{+}\hat{c}_{j}}
\end{equation}%
is a unitary operator.
The matrix elements $C_{kl}$ are
calculated with respect to the states 
\begin{align}
|\alpha \rangle & =|m_{1},m_{2}\ldots ,m_{N}\rangle,  \\
|\beta \rangle & =|n_{1},n_{2},\ldots ,n_{N}\rangle 
\end{align}%
with fermions occupying single particle levels enumerated as $%
m_{1}<m_{2},\ldots ,<m_{N}$ and $n_{1}<n_{2},\ldots ,<n_{N}$
correspondingly, here $N$ is a total number of electrons. We can write these
states as a product of creation operators acting on vacuum 
\begin{align}
|\alpha \rangle & =\hat{c}_{m_{N}}^{+}\hat{c}_{m_{N-1}}^{+}\ldots \hat{c}%
_{m_{1}}^{+}|vac\rangle,  \\
|\beta \rangle & =\hat{c}_{n_{N}}^{+}\hat{c}_{n_{N-1}}^{+}\ldots \hat{c}%
_{n_{1}}^{+}|vac\rangle. 
\end{align}%
The matrix elements ($\ref{matel1}$) can be written as 
\begin{multline}
C_{kl}=\langle vac|\hat{c}_{m_{1}}\hat{c}_{m_{2}}\ldots \hat{c}_{m_{N}}\times  \\
\hat{c}_{k}^{+}\hat{U}\hat{c}_{l}\ \hat{c}_{n_{N}}^{+}\hat{c}_{n_{N-1}}^{+}\ldots 
\hat{c}_{n_{1}}^{+}|vac\rangle 
\end{multline}%
or after commuting the operator $\hat{c}_{k}^{+}$ to the left and $\hat{c}%
_{l}$ to the right we obtain 
\begin{multline}
C_{kl}=(-1)^{p+q}\langle vac|\hat{c}_{m_{1}}\hat{c}_{m_{2}}\ldots \hat{c}%
_{m_{p-1}}\hat{c}_{m_{p+1}}\ldots \hat{c}_{m_{N}}\times  \\
\hat{U}\ \hat{c}_{n_{N}}^{+}\hat{c}_{n_{N-1}}^{+}\ldots \hat{c}_{n_{q+1}}^{+}\hat{c%
}_{n_{q-1}}^{+}\ldots \hat{c}_{n_{1}}^{+}|vac\rangle 
\end{multline}%
where $p$ and $q$ are defined such that $k=m_{p}$ and $l=n_{q}$. Now using
the unitarity of the $\hat{U}$ matrix we can rewrite it as 
\begin{multline}
C_{kl}=(-1)^{p+q}\langle vac|\hat{c}_{m_{1}}\hat{c}_{m_{2}}\ldots \hat{c}%
_{m_{p-1}}\hat{c}_{m_{p+1}}\ldots \hat{c}_{m_{N}}\times  \\
\hat{U}\ \hat{c}_{n_{N}}^{+}\hat{U}^{+}\hat{U}\hat{c}_{n_{N-1}}^{+}\hat{U}%
^{+}\ldots \hat{U}\hat{c}_{n_{q+1}}^{+}\hat{U}^{+} \\
\hat{U}\hat{c}_{n_{q-1}}^{+}\hat{U}^{+}\ldots \hat{U}\hat{c}_{n_{1}}^{+}\hat{%
U}^{+}\hat{U}|vac\rangle 
\end{multline}%
Using the fact that $\hat{U}|vac\rangle =|vac\rangle $ we arrive at the
equation 
\begin{multline}
C_{kl}=(-1)^{p+q}\langle vac|\hat{c}_{m_{1}}\hat{c}_{m_{2}}\ldots \hat{c}%
_{m_{p-1}}\hat{c}_{m_{p+1}}\ldots \hat{c}_{m_{N}}\times   \label{m1} \\
{\tilde{c}}_{n_{N}}^{+}{\tilde{c}}_{n_{N-1}}^{+}\ldots {\tilde{c}}%
_{n_{q+1}}^{+}{\tilde{c}}_{n_{q-1}}^{+}\ldots {\tilde{c}}_{n_{1}}^{+}|vac%
\rangle 
\end{multline}%
where we defined ${\tilde{c}^{+}}_{i}=\hat{U}\hat{c}_{i}^{+}\hat{U}^{+}$.
Applying the Baker-Hausdorff identity we obtain 
\begin{equation}
{\tilde{c}^{+}}_{i}=\sum_{n=0}^{\infty }\frac{i^{n}}{n!}%
M_{ii_{1}}^{T}M_{i_{1}i_{2}}^{T}\ldots
M_{i_{n-1}i_{n}}^{T}c_{i_{n}}^{+}=\sum_{j=0}^{\infty }U_{ij}^{T}c_{j}^{+}
\label{optrans}
\end{equation}%
where matrix $U$ is defined as %an exponential of the matrix $H$
\begin{equation}
U=\mathrm{exp}[iM].
\end{equation}
Substituting Eq. ($\ref{optrans}$) into Eq. ($\ref{m1}$) gives an equation
for the matrix elements 
\begin{equation}\label{matelA}
A_{kl}=\mathrm{D}_{lk}^{-1}\det \mathrm{D}
\end{equation}%
where matrix $D$ is formed from the matrix elements $U_{ij}$ with indices 
$i$ and $j$ spanning the occupied states $m_{1},m_{2},\ldots m_{N}$ and $%
n_{1},n_{2},\ldots n_{N}$ correspondingly.

\section{Numerical evaluation of the correlators}\label{app:num_cor}

In this Appendix we provide details of the numerical procedure, outlined
in Sec. \ref{sub:exact_eval_corr}, which we use at the final stage of
our calculations to evaluate the correlation functions from Eq. (\ref{corr1}).

We consider a system of length $L$ with periodic boundary conditions, which leads to momentum
quantization $k=2\pi n/L $, where $n\in Z$. A cutoff is introduced on the number
of momentum eigenstates, so that $n\in[-N_{max},N_{max}]$, where $N_{max}$ is a positive integer
with the total number of states $N_{tot}=2N_{max}+1$.
The largest value of $N_{max}$ which we used in the calculations in order to check
the convergence was $\sim 1000$, although in most cases $N_{max}\sim500$ was sufficient.
The numerical calculation can be divided into three steps.

{\em Step I.} We generate a matrix $\mathrm{M}$ with dimensions $2{N_{tot}}\times2{N_{tot}}$,
where the factor of two originates from the number of electron channels in the problem. The structure of
this matrix is as follows. In the matrix element $M_{\alpha\beta}$ the indices $\alpha$ and $\beta$
denote channel and momentum of the creation and annihilation operators in the kernel.
We reserve even indices for the first channel and odd for the second channel such that
the matrix element of the kernel operator between states $k$ and $k'$ is represented by $2\times2$ block
in the matrix $\mathrm{M}$. The difference $q=k-k'$ gives the plasmon momentum $q$, which is substituted into
Eq. (\ref{kernel_mom}) with the corresponding form of the plasmon phase shifts. These phase shifts
are obtained from the Eq. (\ref{phase_shifts}) in the case of charging interactions,
Eq. (\ref{ph_sh_exp}) for exponential interactions and can be calculated numerically
using Bogoliubov equations for other interaction potentials as, for example, we have done in Sec.\ref{sub:coulomb-interactions} for Coulomb interactions. For the zero mode i.e. $q=0$ we use Eq. (\ref{phi0}).

We evaluate the kernel from Eq. (\ref{kernel_mom}) for every set of $k,k'$ for each channel (the kernels are different if arm lengths are not equal). Each $2\times2$ block with given $k,k'$ is multiplied by the $2\times2$ matrix $S^{a*}_{1\alpha}S^{a}_{1\beta}$  or $S^{a*}_{2\alpha}S^{a}_{2\beta}$ (according to which channel it originates from), where $\mathcal{S}^{a}$ is defined in the Eq. (\ref{scatt_mat_a}). The difference of the resulting matrices gives the $(k,k')$ block in the matrix $\mathrm{M}$. From this matrix we calculate $\mathrm{H}\equiv \exp(iM)$, which we use in the next step.

{\it Step II.} Next we generate a matrix $\mathrm{D}$ with dimensions $N_p\times N_p$, where $N_p=N_p^{1}+N_p^{2}$ is the total number of particles in both channels, and $N_p^\mu$ is the number in channel $\mu$.  Convergence was achieved in most of the cases for $N_p^{1}\sim N_p^{2}\sim 400$. In the initial state particles occupy momentum eigenstates $-\frac{2\pi}{L}N_{max}\ldots\frac{2\pi}{L}(-N_{max}+N_p^{1,2})$. To study voltage dependence of the correlators we fix number of particles in one channel such that the states $(-\frac{2\pi}{L} N_{max}\ldots 0)$ are occupied and we vary the number of electrons in another channel. The matrix $\mathrm{D}$ is obtained by taking matrix elements of $\mathrm{H}$ which correspond to occupied momentum eigenstates in the initial state.

{\it Step III.} We calculate the determinant and the inverse of the matrix $\mathrm{D}$. Using this and Eq. (\ref{matelA}) we
evaluate correlators $G_{kk'}$ in momentum space for every pair of momentum indices. Correlators in real space are obtained by performing a discrete Fourier transform of $G_{kk'}$.

\end{document}